\documentclass[twocolumn, dvipsnames, twocolappendix]{aastex7}

\usepackage[T1]{fontenc}
\usepackage{amsmath}
\usepackage{textcomp}
\usepackage{gensymb}
\usepackage{graphicx}
\usepackage{url}
\usepackage[caption=false]{subfig}
\usepackage{float} 
\usepackage{soul}
\usepackage{multirow}
\usepackage{newtxmath}

\definecolor{nicegreen}{RGB}{0,180,40}
\hypersetup{colorlinks=True,linkcolor=blue,citecolor=nicegreen,urlcolor=cyan}
\graphicspath{{./}{figures/}}

\newcommand{\noun}[1]{\textsc{\MakeLowercase{#1}}}

\newcommand{\code}{\texttt}

\newcommand{\m}[1]{M$_#1$}
\newcommand{\mwd}{M$_\text{WD}$}
\newcommand{\msun}{$M_\odot$}

\newcommand{\sproc}{\textit{s}-process}

\defcitealias{Rekhi_2024_BaEnrichment}{R24}

\shorttitle{Gaia Barium Dwarfs}
\shortauthors{Rekhi et al.}

\begin{document}

\title{Gaia Barium Dwarfs and Their Ostensibly Ordinary Counterparts}

\correspondingauthor{Param Rekhi}
\email{param.rekhi@weizmann.ac.il}

\author[0009-0001-3501-7852]{Param Rekhi}
\affiliation{Department of Particle Physics and Astrophysics, Weizmann Institute of Science, Rehovot 7610001, Israel}
\email{param.rekhi@weizmann.ac.il}

\author[0000-0001-9298-8068]{Sahar Shahaf}
\affiliation{Max-Planck-Institut f\"ur Astronomie, K\"onigstuhl 17, D-69117 Heidelberg, Germany}
\affiliation{Department of Particle Physics and Astrophysics, Weizmann Institute of Science, Rehovot 7610001, Israel}
\email{sashahaf@mpia.de}

\author[0000-0001-6760-3074]{Sagi Ben-Ami}
\affiliation{Department of Particle Physics and Astrophysics, Weizmann Institute of Science, Rehovot 7610001, Israel}
\email{sagi.ben-ami@weizmann.ac.il}

\author[0000-0002-0430-7793]{Na'ama Hallakoun}
\affiliation{Department of Particle Physics and Astrophysics, Weizmann Institute of Science, Rehovot 7610001, Israel}
\email{naama.hallakoun@weizmann.ac.il}

\author[0000-0001-9590-3170]{Johanna Müller-Horn}
\affiliation{Max-Planck-Institut f\"ur Astronomie, K\"onigstuhl 17, D-69117 Heidelberg, Germany}
\email{mueller-horn@mpia.de}

\author[0000-0002-2998-7940]{Silvia Toonen}
\affiliation{Anton Pannekoek Institute for Astronomy, University of Amsterdam, 1090 GE Amsterdam, The Netherlands}
\email{s.g.m.toonen@uva.nl}

\author[0000-0003-4996-9069]{Hans-Walter Rix}
\affiliation{Max-Planck-Institut f\"ur Astronomie, K\"onigstuhl 17, D-69117 Heidelberg, Germany}
\email{rix@mpia.de}

\begin{abstract}
The recently identified Gaia population of main-sequence--white dwarf (MS+WD) binaries at separations of ${\sim}\,1~{\rm AU}$, often with moderate eccentricities, is not readily reproduced by binary population synthesis models.
Barium stars represent a closely related population whose enrichment in \sproc\ elements both confirms the presence of a WD companion and attests to past binary interaction.
It also indicates that mass transfer occurred at least during the late and post-AGB phases of the WD progenitor, when \sproc\ elements are dredged up.
In this work, we further explore the connection between the astrometrically identified Gaia MS+WD binaries and the classical barium star population.
To achieve this, we used high-resolution FEROS spectroscopy to measure abundances for 30 Gaia DR3 non-single-star binaries, identifying 14 as Ba-enriched. Together with our recent analysis of archival GALAH data, this yields a sample of 40 barium dwarfs with dynamically measured WD masses, compared to only 6 previously known systems with known WD masses at these separations.
We find that, in cases where metallicity is sufficiently low to facilitate efficient \sproc\ production, barium and yttrium enrichment is often detected. This enrichment is also identified in eccentric systems, suggesting that post-AGB mass transfer mechanisms are capable of pumping eccentricity into the orbit or occur without erasing it.
Our results indicate that the Gaia MS+WD binaries trace the population from which barium stars emerge.  Treating the large Gaia-discovered population as an extension of known \sproc\ enriched dwarfs opens an avenue to empirically constrain their formation and evolution.
\end{abstract}

\section{Introduction}

Barium stars \citep{Bidelman_1951_BaII} were first discovered as red-giant stars of spectral type G-K with enhanced abundances of elements produced by the slow neutron capture process (\sproc). Barium dwarfs are main-sequence (MS) stars of classes F-K with similar \sproc\ element enhancement \citep{North_1994_NatureSTR}. This type of chemical enrichment has long been interpreted as a fossil record of past binary interactions. 

Since \sproc\ elements cannot be produced internally by stars before the asymptotic giant branch (AGB) phase, their presence is understood as a signature of mass accretion from an AGB companion that had since evolved into a white dwarf \citep[WD; see, e.g.,][and references therein]{Jorissen_2019_BariumRelated}. 
During thermal pulses in the late-AGB phase, \sproc\ elements such as barium and yttrium are mixed to the stellar surface and, if the AGB star fills its Roche Lobe, can be transferred to a binary companion. The detectability of these elements in the accretor’s photosphere depends on both the total amount of material accreted and the degree of dilution in the stellar envelope \citep{Husti_2009_BariumStars}. Surface \sproc\ enhancement, therefore, offers evidence of mass transfer, independent of the system's current orbital configuration.

The third data release (DR3) of Gaia \citep{GaiaCollaboration_2023_GaiaData} includes a catalog of non-single-star solutions \citep[NSS;][]{GaiaCollaboration_2023_NSS} for unresolved binaries, enabling astrometric orbit characterization across a wide range of orbital separations. \citet{Shahaf_2024_TriageGaia} identified ${\sim}\,3200$ MS+WD astrometric binaries in the NSS catalog, in which a faint WD accompanies a luminous MS star. These binaries have periods between $100-1000$~days (corresponding to orbital separations of ${\sim}\,1$\,AU), and a substantial fraction of them exhibit moderate to high eccentricities ($e$ up to 0.7). Standard binary population synthesis models do not predict this population, as they preferentially produce systems with shorter or longer orbital periods (see figure 16 of \citealt{Shahaf_2024_TriageGaia} and discussion thereof). Its empirical properties and the processes that enabled it are the subject of ongoing studies \citep[e.g.,][]{Hallakoun_2024_DeficitMassive, Yamaguchi_2024_WidePostcommon, Yamaguchi_2025_PopulationDemographics, Ironi_2025_InitialtofinalMass, Rubio_2025_CalibrationBinary}.

Identifying \sproc\ enrichment in even part of this sample would link these systems to Barium stars, partially constrain the timing of mass transfer to the late- and post-AGB phase, and place them in a broader evolutionary context. To search for such signatures, \citet[][\citetalias{Rekhi_2024_BaEnrichment} hereafter]{Rekhi_2024_BaEnrichment} combined this Gaia-selected sample with GALAH DR3 spectroscopy. They measured \sproc\ abundances in 102 MS+WD binaries, and found that barium and yttrium enrichment in the MS primaries correlates with WD mass and metallicity. However, they did not find significant link with the orbital properties of the systems, possibly due to the limited sample size.

This study aims to extend the work of \citetalias{Rekhi_2024_BaEnrichment} by targeting eccentric MS+WD systems in the 100-1000\,d period range. To achieve this, we conduct spectroscopic analysis to measure \sproc\ abundances of 21 eccentric MS+WD binaries, complemented by a comparison sample of 12 systems with WD masses above 0.75\,{\msun} with nearly circular orbits. The targets were observed using the Fiber-fed Extended Range Optical Spectrograph \citep[FEROS;][]{Kaufer_1998_FEROSNew, Kaufer_1999_CommissioningFEROS}, mounted on the MPG/ESO 2.2-meter telescope at the La Silla Observatory in Chile. FEROS covers the spectral range from 350 to 920\,nm at a resolving power of $R \sim 48{,}000$, making it well-suited for precise measurements of chemical abundances.

Section~\ref{sec: sample and obs} describes the sample selection and spectroscopic campaign. Section \ref{sec:chem abd procedure} details our procedure for estimating stellar parameters and Ba and Y abundances. We discuss our results and their astrophysical implications in Section \ref{sec: res and disc}.
Section \ref{sec: outlook} provides an outlook for future work.

\section{Sample Selection and Observations} \label{sec: sample and obs}
\subsection{Sample Selection}

\begin{figure}
\centering
\includegraphics[width=\columnwidth]{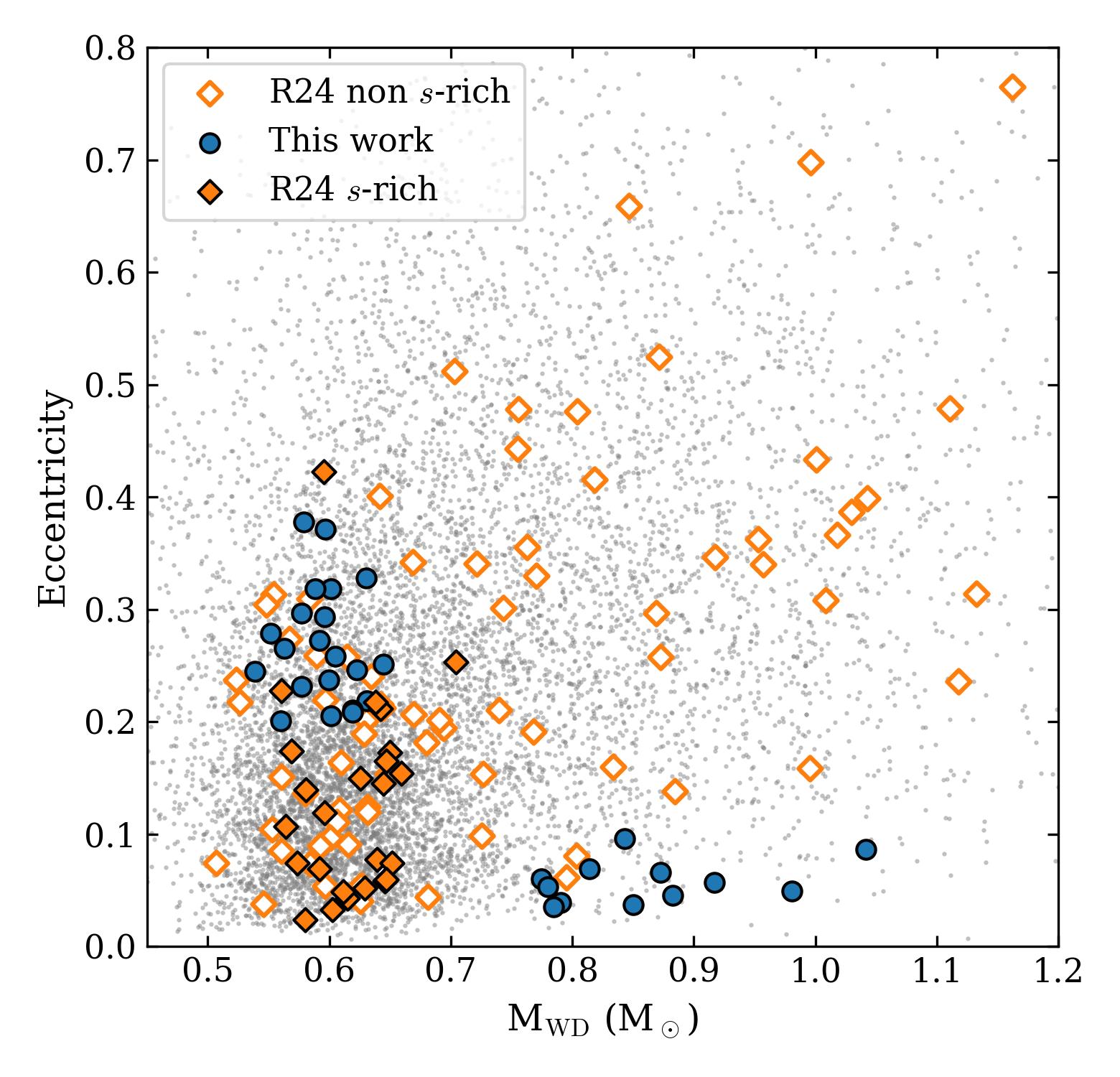}
\label{fig:feros sample eM}
\caption{Eccentricity-\mwd\ diagram of the FEROS sample (blue). The sample includes 21 systems with $e>0.2$ and \mwd\ $<0.65$\,\msun, complemented with a comparison sample of 12 systems with $e<0.1$ and \mwd\ $>0.75$~\msun. These systems probe regions of parameter space largely unexplored by \sproc-enhanced binaries in the GALAH sample of \citetalias{Rekhi_2024_BaEnrichment} (orange). The \citet{Shahaf_2024_TriageGaia} sample is shown in grey for reference.}
\end{figure}

Our sample is drawn from the ${\sim}\,3200$ MS+WD candidates identified by \citet{Shahaf_2024_TriageGaia} using the astrometric triage method \citep[see][]{Shahaf_2019_TriageAstrometric, Shahaf_2023_TriageGaia}. This method uses the astrometric mass-ratio function (AMRF) to link the photocentric orbit to the binary's mass and flux ratios. Systems with AMRF values exceeding the maximum allowed by MS mass–luminosity relations cannot host a single MS companion and are classified as \textit{non-class-I} binaries. The faint companions in these systems are likely WDs or close MS binaries. To reduce contamination from the latter, \citet{Shahaf_2024_TriageGaia} excluded systems with infrared excess. This process produced a sample of highly probable MS+WDs. The purity of this sample has been validated by several studies \citep[e.g.,][]{Yamaguchi_2024_WidePostcommon, Yamaguchi_2025_PopulationDemographics}.

We selected two subsamples from the AMRF catalog for spectroscopic follow-up (Figure~\ref{fig:feros sample eM}). The first consists of systems with WD masses of $0.5$–$0.65$\,\msun\ and eccentricities $e>0.2$. This mass range corresponds to the regime identified by \citetalias{Rekhi_2024_BaEnrichment} as favorable for \sproc\ production during the AGB phase. Detecting enrichment in these systems would indicate past mass transfer when the progenitor radius was comparable to the current orbital separation, providing a test of whether high eccentricity can survive binary interaction or arise afterward. This region of parameter space was sparsely sampled in \citetalias{Rekhi_2024_BaEnrichment}. This work extends it by targeting additional eccentric systems.

The second subsample is comprised from systems with WD more massive than $0.75$\,\msun\ and eccentricities below $0.1$. These likely descend from progenitors with masses $>3$\,\msun, which are expected to produce lower intrinsic \sproc\ yields \citep{Karakas_2016_StellarYields}. This high-\mwd, low-eccentricity regime is absent from the \citetalias{Rekhi_2024_BaEnrichment} sample, and complementary in its properties to the first one. 

Our targets were selected to be observable from La Silla, Chile, in January 2025, at airmass below 1.8 for sufficient time to reach predicted signal-to-noise ratios (SNRs) of $\sim 50$ per pixel at 4500\,\AA. We excluded systems already observed by GALAH and limited the sample to $B < 15$\,mag. These criteria respectively yielded 32 and 16 viable targets out of 529 and 41 total objects in the two categories described above, of which we selected the brightest 21 and 12 targets.
6 targets from the high-eccentricity subsample were also observed in an additional epoch in March 2025.
Similar to the \citetalias{Rekhi_2024_BaEnrichment} sample, the systems in this work have near-solar primary masses, ranging between 0.8 to 1.2 \msun.

\subsection{Spectroscopic observations}

Each target was observed with FEROS in $2-3$ consecutive exposures to assist in cosmic-ray removal; the exposures were co-added before analysis. The spectra were reduced using the \noun{CERES} pipeline \citep{Brahm_2017_CERESSet}, which provides continuum-normalized spectra. 
We masked normalized-flux values exceeding 1.2 or less than $-0.2$ in order to avoid artifacts in the spectra.
Due to particularly low SNR in our spectra below 4500\,\AA, this region was excluded from the analysis. 

The target exposures were set to obtain an SNR per pixel of ${\sim}\,50$ at 4500\,\AA\ based on the FEROS exposure time calculator\footnote{\url{https://www.eso.org/observing/etc/bin/gen/form?INS.NAME=FEROS+INS.MODE=spectro}} (SNRs at longer wavelengths would be higher in FEROS spectra). However, the observed spectra had significantly lower SNRs around 4500\,\AA\, between 3.1 and 8.5, with a median SNR of 4.4. Nevertheless, most spectra had sufficient signal to identify the Fe and Ba lines required for our analysis. An example of a recorded spectrum with a typical SNR at the vicinity of the spectral lines analyzed in this work is shown in Figure~\ref{fig:sample fit idx7}.

\section{Chemical abundances} \label{sec:chem abd procedure}

\begin{figure*}
\centering
\subfloat{
    \includegraphics[width=1.05\columnwidth]{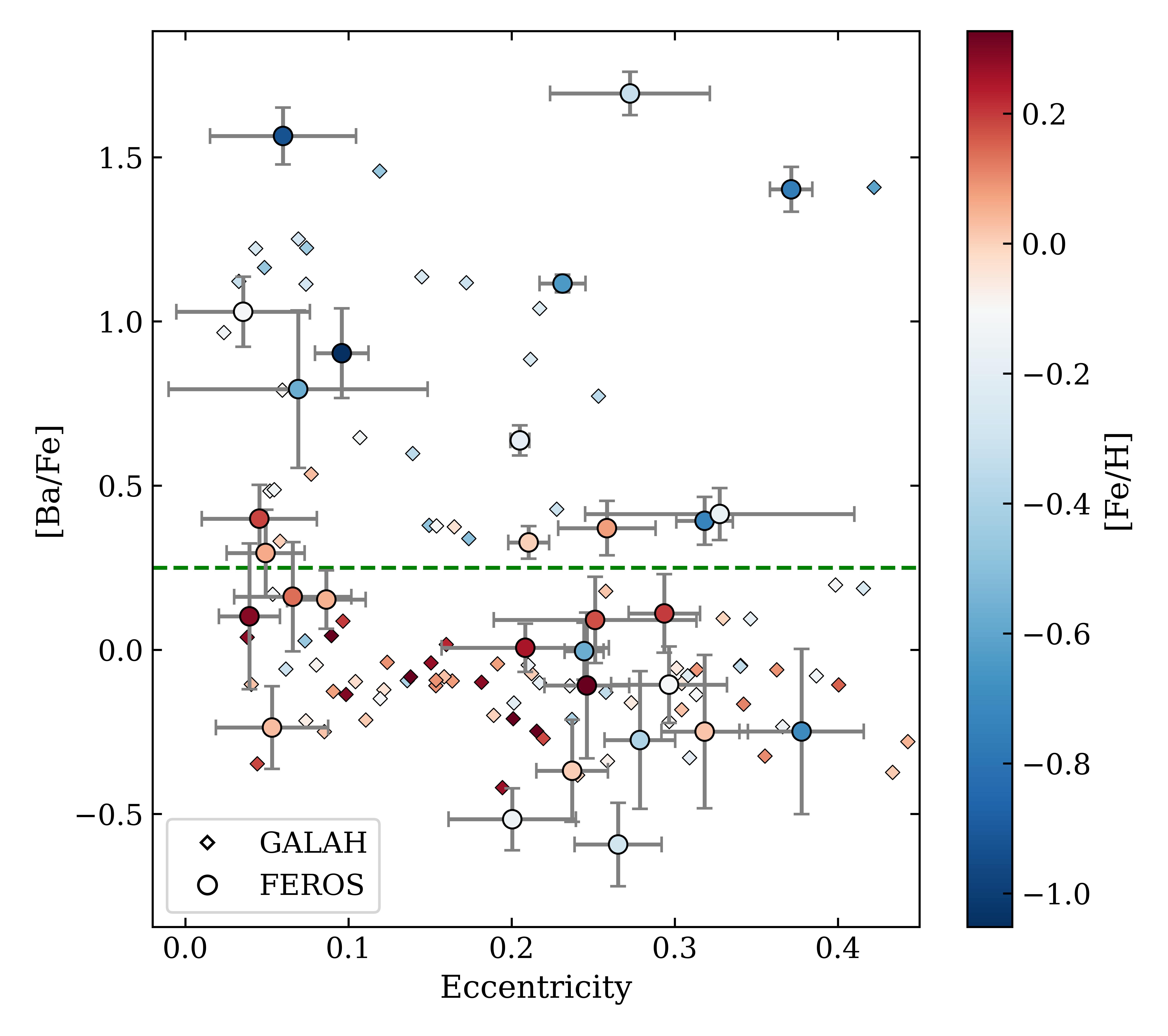}
    \label{fig: Ba ecc}
}
\subfloat{
    \includegraphics[width=1.05\columnwidth]{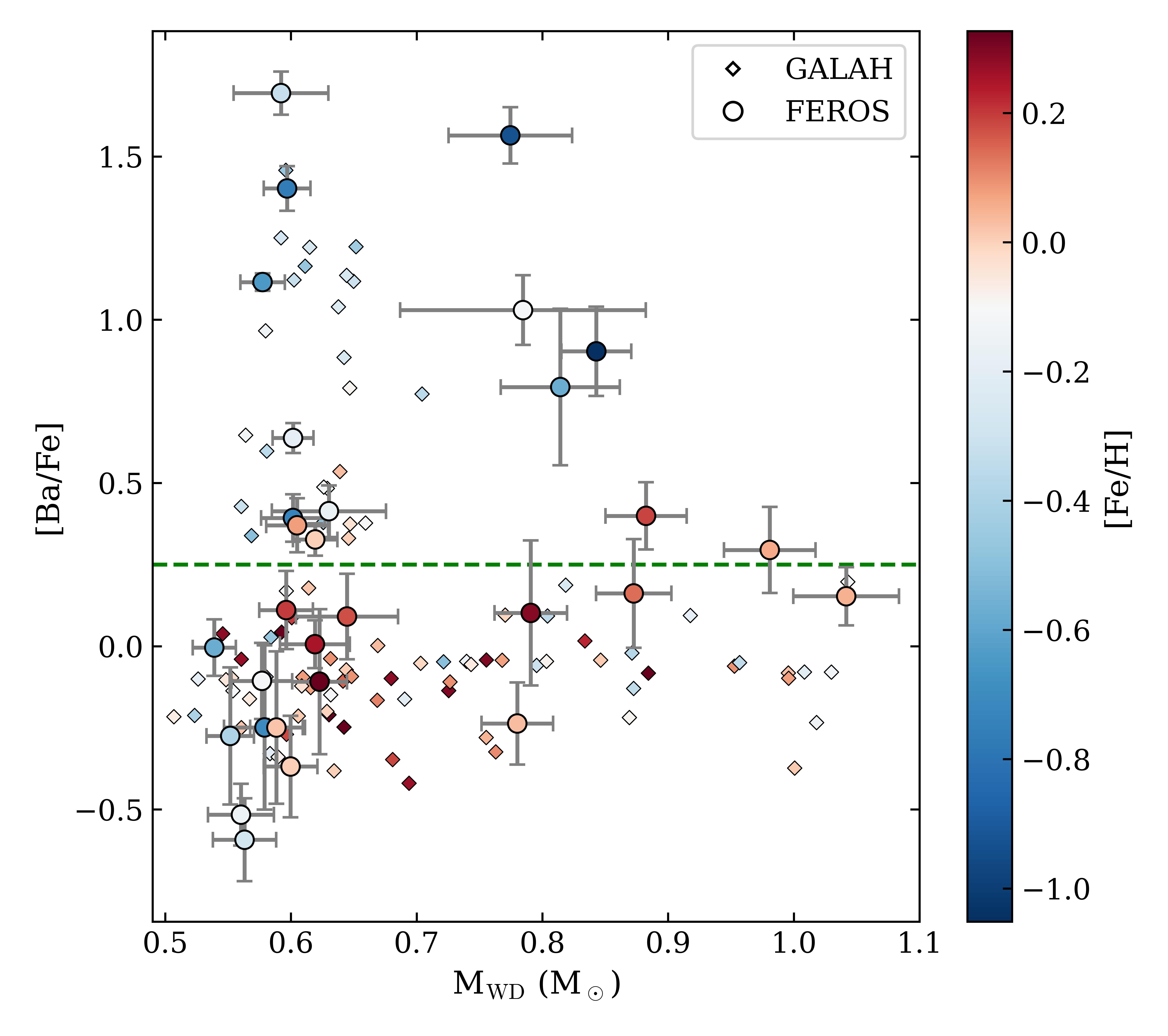}
    \label{fig: Ba mwd}
}
\caption{[Ba/Fe] as a function of (a) eccentricity and (b) \mwd, colored by [Fe/H]. The GALAH sample from \citetalias{Rekhi_2024_BaEnrichment} is also plotted for reference. The green dashed line denotes the boundary beyond which Ba is enhanced. We detect 8 Ba-enhanced systems at $e>0.2$, doubling the \citetalias{Rekhi_2024_BaEnrichment} eccentric \sproc-enriched sample. However, an additional 6-8 eccentric systems having commensurate metallicities are not \sproc\ enriched, indicating unknown factors influencing mass-transfer in eccentric systems. We also identify 6 Ba-enhanced systems hosting massive WDs (\mwd~$\in [0.75,0.1]$\,\msun) at low eccentricities, demonstrating that enrichment can occur even where \sproc\ yields are expected to be suppressed due to their relatively high progenitor masses. [Y/Fe] plots are shown in Appendix \ref{app:addn plots}, demonstrating that Y and Ba enhancements occur in tandem.}
\label{fig: Ba color plots}
\end{figure*}

The spectral modeling for fitting the metallicity and \sproc\ element abundances was carried out using \noun{pymse} \citep[][v0.4.188]{Piskunov_2017_SpectroscopyMade, Wehrhahn_2023_PySMESpectroscopy}. Spectral line data was obtained from the VALD3 database\footnote{\url{https://vald.astro.uu.se/}} \citep{Kupka_1999_VALD2Progress, Piskunov_1995_VALDVienna, Ryabchikova_2015_MajorUpgrade}, with an atomic and molecular line list generated for a Sun-like star that included all lines with central depths $\geq 0.001$. Using NLTE departure coefficient grids from \citet{Amarsi_2020_GALAHSurvey}, \noun{pysme} supports NLTE synthesis for Fe and Ba, however analysis for Y assumed LTE. To account for telluric features in the observed spectra, we employed \noun{TelFit} \citep[][v1.4.0]{Gullikson_2014_CorrectingTelluric}, generating synthetic telluric models based on site-specific observational conditions, broadened to $R=48,000$. Spectral regions where the modeled telluric transmission fell below 0.97 were masked from analysis.

We derived stellar parameters ($T_\text{eff}$, [Fe/H]) and line-broadening velocities ($v_\text{broad}$, $v_\text{mic}$) from carefully selected Fe lines. Surface gravities ($\log g$) were fixed to values from Gaia XP–based catalogs \citep{Andrae_2023_RobustDatadriven,Zhang_2023_Parameters220}, as the chosen lines are minimally pressure-broadened and therefore insensitive to $\log g$. For each star, spectral fits were obtained by performing MCMC fitting of \noun{pysme} models to these Fe lines, with local RV estimation and continuum detrending performed for every line segment. 
A full description of the Fe-line selection, continuum treatment, RV corrections, and spectral-fitting procedures is provided in Appendix~\ref{app:rv_feh_vbroad}.

Ba and Y abundances were then measured using an analogous MCMC procedure, fixing the stellar parameters to the previously determined values and applying the same RV and continuum treatment. We adopted three Ba lines (4554.03, 5853.68, 6496.90 \AA\ in air) and three Y lines (4900.12, 5087.42, 5200.41 \AA\ in air), chosen for their depth and lack of blending across the relevant stellar parameter space. For the six two-epoch targets, stellar parameters and abundances were inferred in a joint MCMC framework, with each spectrum independently preprocessed.

Because of the low SNR of our spectra, we excluded targets without clearly distinguishable Ba and Y lines from the corresponding analysis. We obtained Fe and Ba abundances for 30 stars: 20 in the ($e>0.2$, \mwd\,$\in\,[0.5,0.65]$\,\msun) subsample and 10 in the (\mwd\,$>0.75$\,\msun, $e<0.1$) subsample. Of these, Y abundances could be measured for only 7 and 3 stars, respectively, as the Y lines are considerably shallower than the Ba lines at comparable abundance levels. All fitted parameters and their uncertainties are listed in Table \ref{tab:data_table}. Following \citet{VanDerSwaelmen_2017_MassratioEccentricity} and \citetalias{Rekhi_2024_BaEnrichment}, we define enrichment as [Ba/Fe] or [Y/Fe] exceeding 0.25 dex. A comparison of [Ba/Fe] and [Y/Fe] abundances is shown in Figure \ref{fig:ba y comp}, confirming that Ba and Y enhancement typically occur together.

\section{Results and Discussion} \label{sec: res and disc}

Figure \ref{fig: Ba color plots} presents our measured Ba abundances as functions of eccentricity, WD mass and metallicity, with corresponding Y abundance plots shown in the appendix (Figure \ref{fig: Y color plots}). 
As demonstrated in \citetalias{Rekhi_2024_BaEnrichment}, as well previous Ba star studies \citep[e.g.,][]{Jorissen_2019_BariumRelated, Roriz_2021_HeavyElements, Vilagos_2024_BariumStars}, the detection of \sproc-enhancement depends strongly on metallicity and WD mass, which govern the efficiency of \sproc\ nucleosynthesis in the WD progenitor \citep[see ][]{Busso_1999_NucleosynthesisAsymptotic, Busso_2001_NucleosynthesisMixing, Karakas_2016_StellarYields}. Our results align with this overarching paradigm while additionally highlighting the influence of eccentricity, as elaborated below.

\subsection{AMRF MS+WD Systems as Ba-Stars}

\begin{figure}
    \centering
    \includegraphics[width=0.95\columnwidth]{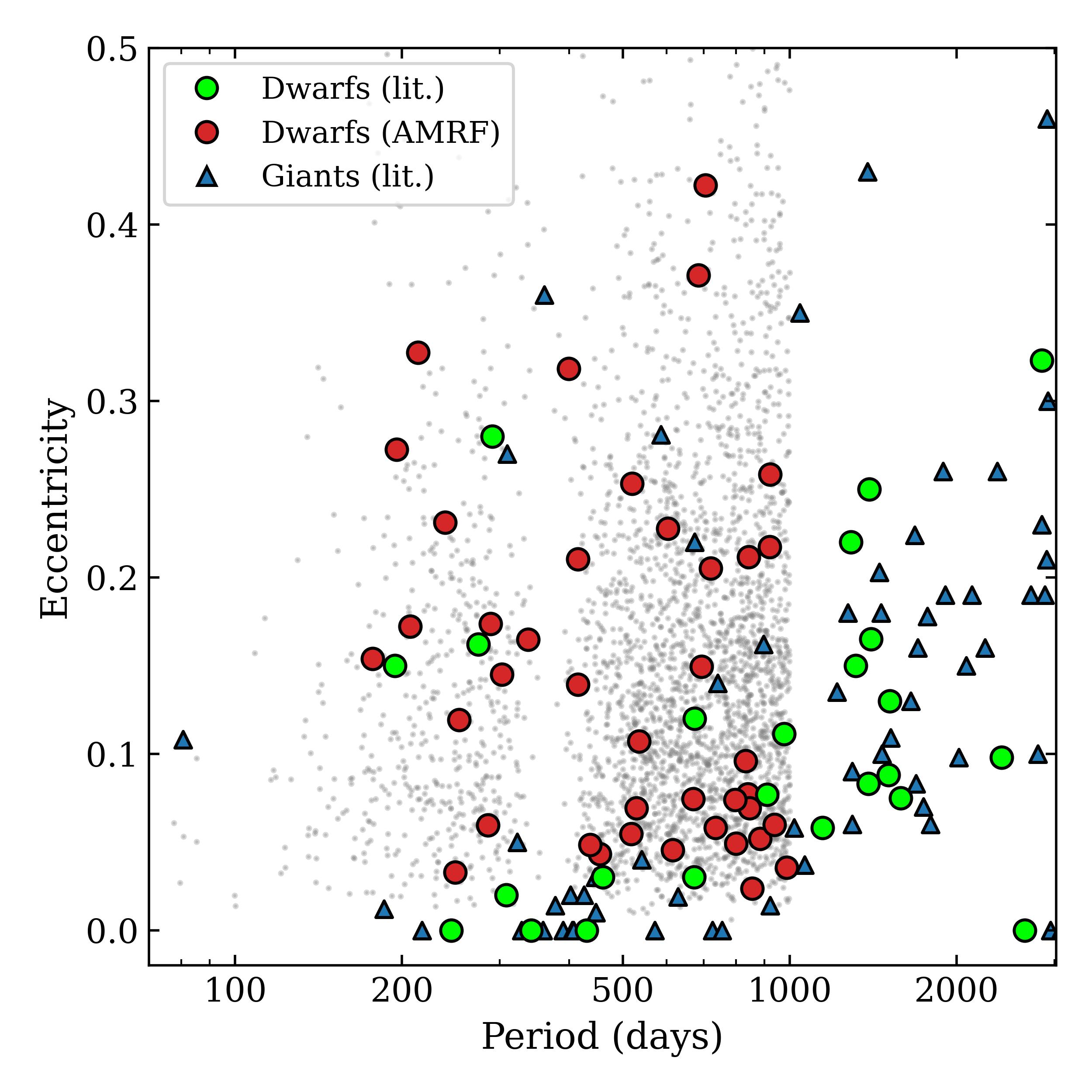}
    \caption{Period-Eccentricity diagram of the \citet{Shahaf_2024_TriageGaia} AMRF sample (gray), its Ba-enriched subset from this work and \citetalias{Rekhi_2024_BaEnrichment} (red) and Ba/CH/CEMP-s stars from literature \citep[green/blue; ][]{Jorissen_2016_BinaryProperties, Escorza_2019_BariumRelated, Escorza_2023_BariumRelated}. Our samples uncover a significantly larger population of Ba dwarfs at $e$ > 0.1 at P < 1000\,d, which is underrepresented in literature. The AMRF sample thus provides a systematic route to identifying large samples of Ba dwarfs with orbital parameters and companion WD masses from the Gaia NSS catalog.}
    \label{fig:P e general}
\end{figure}

\begin{figure}
    \centering
    \includegraphics[width=0.98\columnwidth]{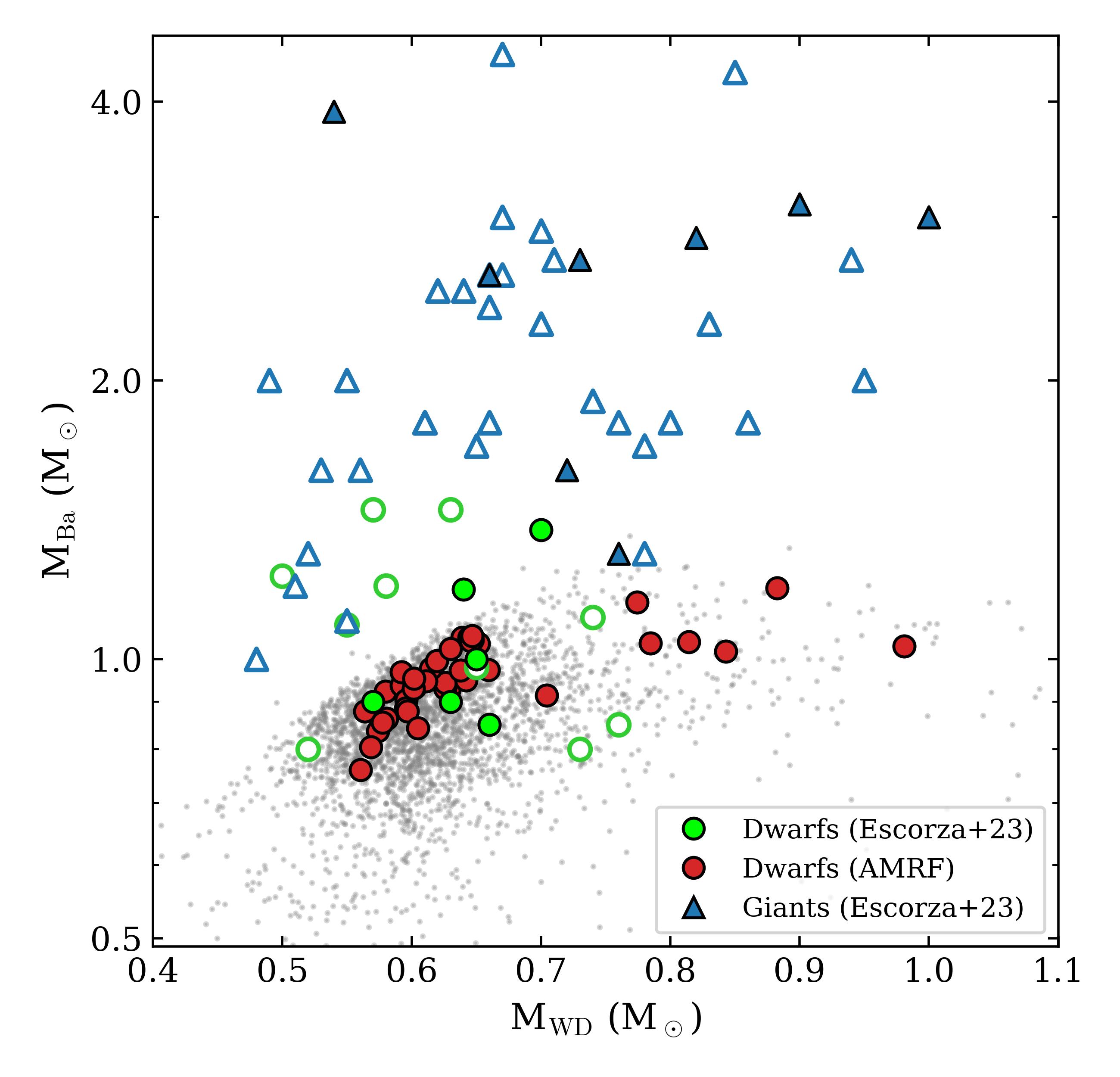}
    \caption{Primary and WD masses of the Ba star samples from Figure \ref{fig:P e general}. Systems with P > 1000\,d are plotted as open circles.
    The WD companions of Ba dwarfs have masses around 0.6\,\msun\ in both samples, indicating progenitor core development similar to field WDs. The generally higher WD masses in the Ba giants sample likely reflect a detection bias, as these systems tend towards higher primary masses.}
    \label{fig:mwd m1}
\end{figure}

The results of \citetalias{Rekhi_2024_BaEnrichment} and this work show that, whenever metallicity permits, a substantial fraction of MS primaries in the AMRF sample exhibit Ba enhancement ([Ba/Fe]\,$>0.25$\,dex), consistent with expectations from AGB \sproc\ yields.

Figure~\ref{fig:P e general} compares the period-eccentricity distribution of the AMRF sample with literature samples of Ba dwarfs and giants \citep{Escorza_2019_BariumRelated, Escorza_2023_BariumRelated} and analogous CH/CEMP-s stars \citep{Jorissen_2016_BinaryProperties}. Both the overall AMRF distribution and the Ba-enhanced subset align with these populations, while extending coverage of Ba dwarfs to periods less than 1000 days (many with $e>0.1$), a region underrepresented in previous work. 
At these separations, the inflated radii of giant stars raise the possibility of dynamical interactions with their white dwarf companions. Although the current samples are too limited to draw statistical conclusions, the forthcoming Gaia DR4 release is expected to provide a larger sample of astrometric giant–white dwarf systems, enabling a more robust investigation of this effect.

Figure~\ref{fig:mwd m1} further shows similar distribution of estimated WD masses (with MS companions) between the AMRF and the RV-based sample of \citet{Escorza_2023_BariumRelated}. The WD companions of Ba dwarfs have masses around 0.6\,\msun\ in both samples, indicating progenitor core development similar to field WDs. The generally higher WD masses in the Ba giants sample likely reflect a detection bias, as these systems tend towards higher primary masses.

\begin{figure}
    \centering
    \includegraphics[width=\columnwidth]{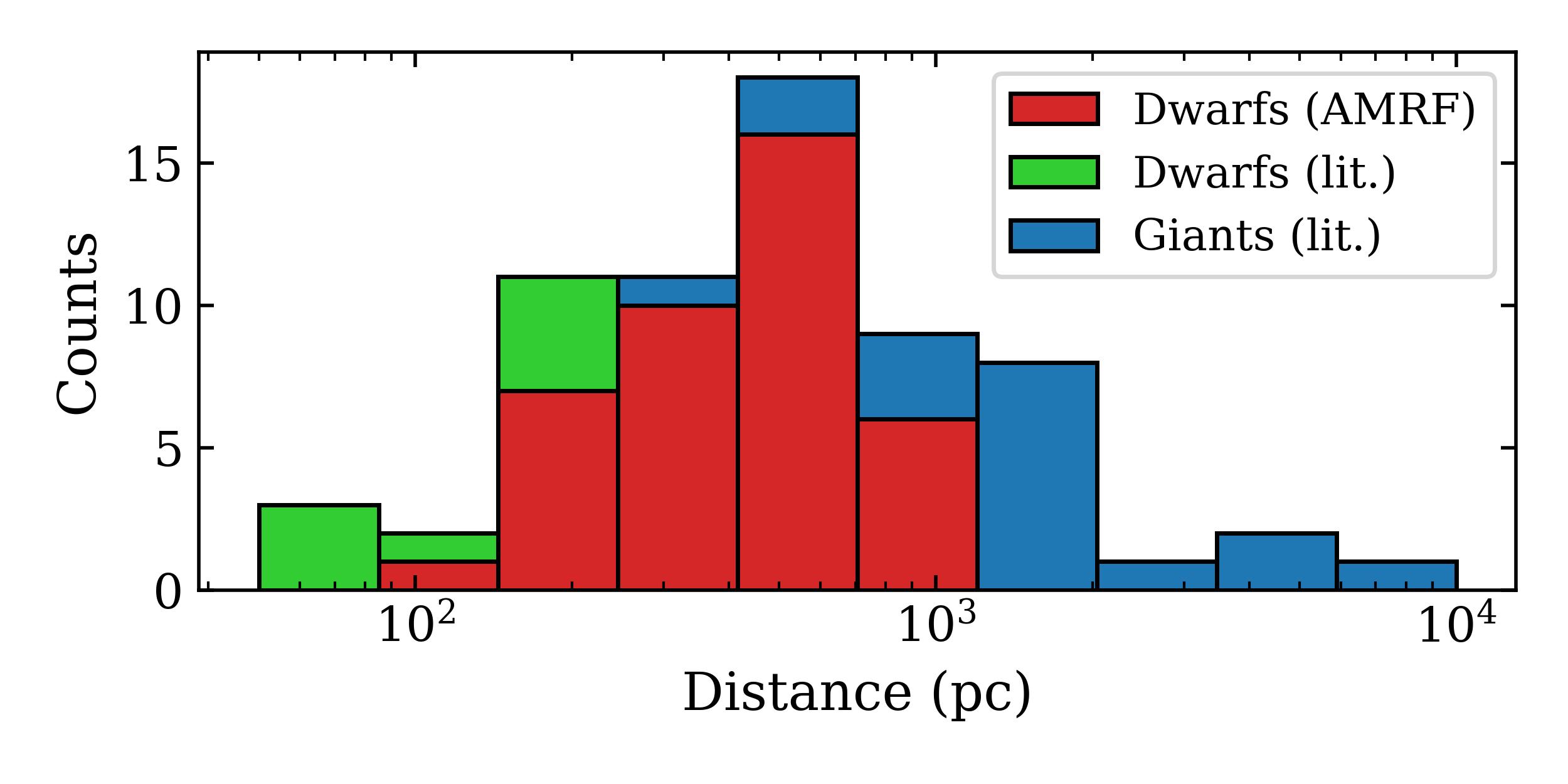}
    \caption{Stacked distance histogram of the Ba star samples from Figure \ref{fig:P e general} having P < 1000\,d. The AMRF method increases the volume probed for Ba Dwarfs by an order of magnitude.}
    \label{fig:dist hist}
\end{figure}

These results indicate that the Gaia AMRF MS+WD binaries are products of interaction with AGB stars, likely through the same channels that produce Ba/CH/CEMP-s systems. Depending on metallicity, orbital period, and selection effects, they may manifest as classical Ba stars. Gaia instead probes a complementary regime of shorter periods and a wider metallicity range, where enrichment is often weaker or harder to detect. Thus, even if not spectroscopically classified as \sproc-enhanced, these systems remain analogous in their overall characteristics. Notably, the current AMRF sample probes a volume an order of magnitude larger than the literature populations (Figure~\ref{fig:dist hist}), providing a far broader view of the underlying parameter space. Future Gaia releases, extending to longer periods, will establish the AMRF sample as a benchmark for modeling binary interaction in intermediate-mass systems.

\subsection{Eccentricity Dependence of Mass Transfer}

In our sample of MS+WD binaries, both in this work as well as \citetalias{Rekhi_2024_BaEnrichment}, we detect pronounced Ba and Y overabundance ([Ba/Fe], [Y/Fe] > 0.25) in many MS primaries up to the maximal orbital eccentricities probed ($\approx 0.4$; Figures~\ref{fig: Ba ecc} and \ref{fig: Y ecc}). Such \sproc\ enrichment clearly indicates past mass transfer---at least in the late and post-AGB phase---from the WD progenitor to the companion MS star; however, the persistence of significant eccentricity in these shows that tidal circularization, expected to accompany such mass transfer, either was much weaker than expected, or did not fully erase the orbital eccentricity.

However, this Ba enhancement seems to be sporadic, as only around half the systems with high eccentricity and sufficiently low metallicity demonstrate it. This points to as yet unconfirmed factors influencing mass-transfer in eccentric systems. One possibility is that these systems had different initial orbital parameters, perhaps being significantly more eccentric or more widely separated, reducing their accretion efficiency. Another possibility is that the progenitors of these WDs were much more massive than expected from the canonical single star initial-final mass ratio functions \citep{Cummings_2018_WhiteDwarf, Cunningham_2024_InitialfinalMass}, and had correspondingly lower \sproc\ yields. Readers can refer to \citet{Ironi_2025_InitialtofinalMass} and \citet{Shahaf_2025_WhiteDwarf} for a detailed discussion.

We also identify high Ba enhancement in six near-circular systems hosting massive WDs (\mwd~$\in [0.75,0.1]$\,\msun; Figure~\ref{fig: Ba mwd}), demonstrating that enrichment can occur even where \sproc\ yields are expected to be suppressed due to their relatively high progenitor masses. The presence of Ba enhancement in these systems likely reflects an interplay between three factors: suppression of \sproc\ yields with high progenitor masses, increased yields at low metallicities, and potentially enhanced accretion efficiency associated with low eccentricity. Although the current data are too sparse to disentangle these effects, these systems extend the empirical evidence for mass transfer to encompass higher-mass WDs.

\subsection{Formation Channels}

Our results add to a growing body of work documenting significant eccentricities in various classes of post-interaction binaries \citep[e.g., ][]{Karakas_2000_EccentricitiesBarium, Jorissen_2016_BinaryProperties, Jorissen_2019_BariumRelated, Vos_2017_OrbitsSubdwarfB, VanDerSwaelmen_2017_MassratioEccentricity,Oomen_2018_OrbitalProperties, Muller-Horn_2025_HIP15429}. A particularly relevant parallel is provided by \citet{Oomen_2018_OrbitalProperties}, who studied a sample of 33 post-AGB+MS binaries with orbital periods of $100-3000$~days, most of which exhibit significant eccentricity ($> 0.1$). Our work demonstrates that these eccentricities persist through the post-AGB phase to the WD formation, diminishing the possibility that the Oomen at al. systems could still circularize via accretion and/or interaction with the post-AGB planetary nebula or circumbinary (CB) disk.
Tidal evolution theory predicts that binaries in which one component fills its Roche lobe should circularize on relatively short timescales, far shorter than the expected duration of the mass transfer phase in the progenitor system \citep[see, e.g., ][]{PortegiesZwart_1996_PopulationSynthesis, Nelemans_2001_PopulationSynthesis, Sepinsky_2007_InteractingBinaries, Sepinsky_2009_InteractingBinaries, Toonen_2012_SupernovaType, Toonen_2013_EffectCommonenvelope}. Multiple models have been proposed to explain this seemingly anomalous population of eccentric post-interaction systems. Broadly, these fall into two categories:
(1) Mechanisms that avoid strong tidal circularization altogether;
(2) Mechanisms that actively pump eccentricity, counteracting canonical tidal damping. 

In the first category, one possibility is that tidal forces are much weaker than expected in certain systems \citep[e.g., ][]{Nie_2017_OrbitalNature}, allowing the Roche-lobe overflow (RLOF) phase to conclude before circularization can occur. 
This seems unlikely for our sample, except perhaps in extremely eccentric cases, because in most systems at these separations, the WD progenitor would have substantially filled its Roche lobe for much of the RGB and the entirety of the AGB phase, thereby experiencing significant tidal forces \citep{Zahn_1977_TidalFriction}. Another possibility is that mass transfer proceeds primarily via stellar winds, rather than RLOF, which can produce non-trivial eccentricity evolution \citep{Boffin_1988_CanBarium, Boffin_2015_MassTransfer, Saladino_2018_GoneWind, Saladino_2019_EccentricBehaviour, Schroder_2021_EvolutionBinaries}, though the former objection may also apply here. A third option is that the observed eccentricity reflects the presence of a tertiary companion, either through direct orbital perturbations or because the tertiary itself was the AGB donor, later being ejected or migrated to a wider orbit after the interaction \citep{Perets_2012_TripleEvolution}.

To investigate the third possibility, we employed the algorithm of \citet{El-Badry_2021_MillionBinaries} to search for wide companions to our systems using their NSS-derived astrometric parameters.
No tertiary matches were found for any of the \sproc-enriched systems reported here or in \citetalias{Rekhi_2024_BaEnrichment}. However, the detection probability declines sharply for companions fainter than $G \sim 19$, which, for our sample, effectively excludes all WDs with cooling ages older than a Gyr as well as the majority of M dwarfs. A more definitive assessment of this scenario will thus require deeper surveys, such as future Gaia data releases and surveys by the Vera Rubin Observatory \citep{Ivezic_2019_LSSTScience}.

In the second category, a frequently invoked eccentricity-pumping mechanism is the formation of a CB disk from AGB ejecta, with resonant disk–binary interactions driving eccentricity growth \citep{Dermine_2013_EccentricitypumpingPostAGB, Vos_2015_TestingEccentricity, DOrazio_2021_OrbitalEvolution, Siwek_2023_OrbitalEvolution, Valli_2024_LongtermEvolution}. Other proposed mechanisms include phase-dependent mass loss via RLOF and/or winds \citep{Dosopoulou_2016_OrbitalEvolution, Parkosidis_2025_RethinkingMass}, and tidally enhanced mass loss at periastron \citep{BonacicMarinovic_2008_OrbitalEccentricities, Kashi_2018_CounteractingTidal}. As demonstrated by \citet{Parkosidis_2025_RethinkingMass}, phase-dependent RLOF mass-loss can cause eccentricity-pumping, but comprehensive population synthesis modeling is required to determine whether they can reliably reproduce the orbital characteristics presented here and in other studies. 

However, we note that many existing studies often yield contradictory results or fail to explore the full parameter space relevant to our systems, and those that especially explore the $100-1000$\,d period range covered by our sample have been unable to reproduce the observed eccentricities \citep{Rafikov_2016_ECCENTRICITYEXCITATION, Saladino_2019_EccentricBehaviour, Oomen_2020_DiscbinaryInteractions, Krynski_2025_FormationBa}.
The recent work by \citet{Krynski_2025_FormationBa} presented detailed models of the evolution of AGB+MS binaries, incorporating many of the mechanisms discussed above, including classical and wind RLOF, tidally enhanced winds, non-conservative mass transfer, and eccentricity pumping due to a CB disk. Even under favorable assumptions, however, the models failed to reproduce the observed eccentricities of post-interaction MS+WD binaries with periods shorter than $\sim2000$~days. This discrepancy indicates that key aspects of the underlying processes remain poorly understood.

\section{Outlook} \label{sec: outlook}

\begin{figure}
    \centering
\includegraphics[width=\columnwidth]{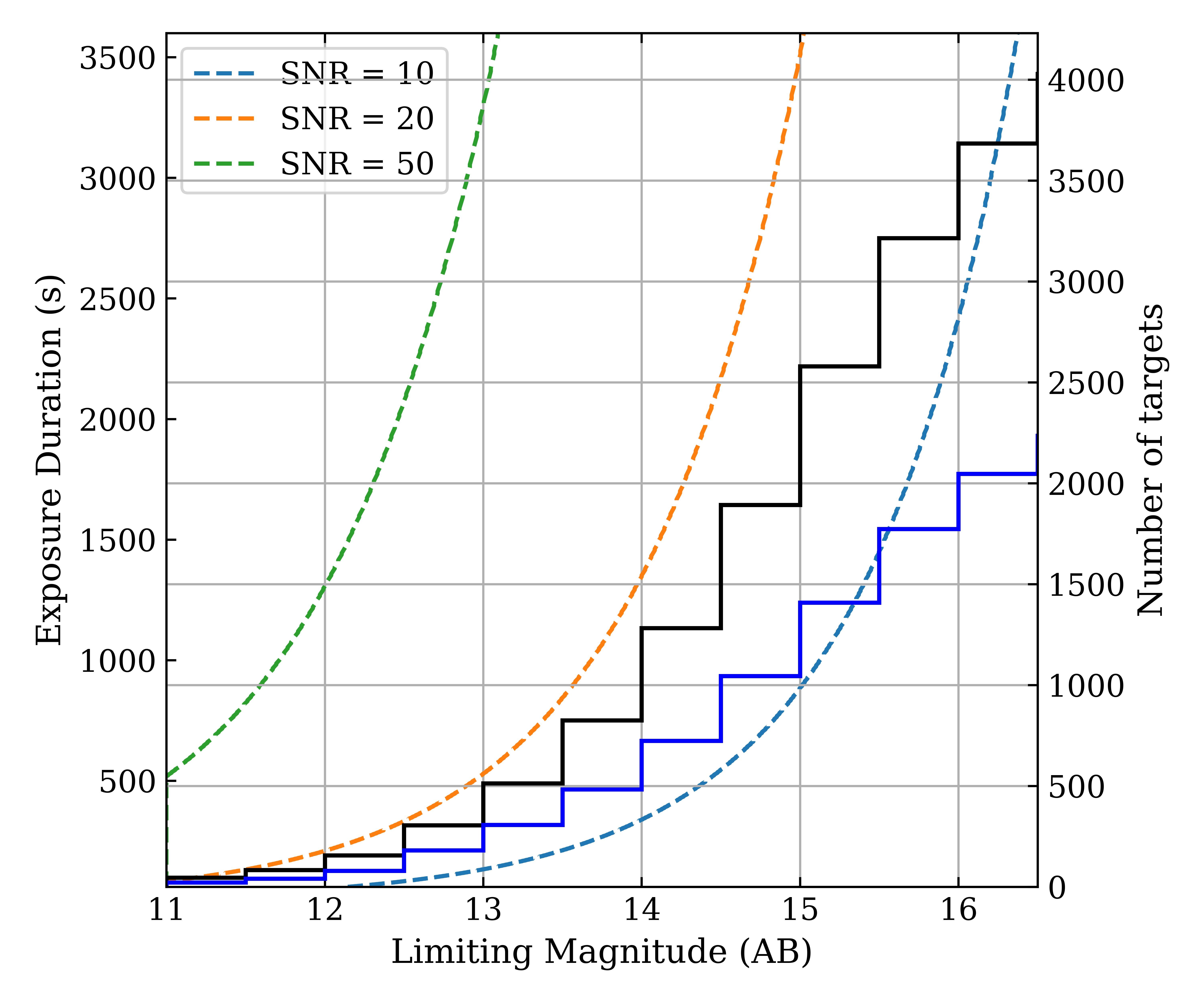}
    \caption{MAST-HighSpec limiting magnitudes for SNRs 10, 20 and 50 as a function of exposure duration. The cumulative distribution of MS+WD systems from \citet{Shahaf_2024_TriageGaia} having [Fe/H] $<0$ observable from Israel is plotted in black. The subset of these with $e>0.2$ is plotted in blue. With 20 MAST telescopes operating simultaneously with HighSpec, we can observe $\sim 1000$ of the brightest targets (including half with $e > 0.2$) at SNR > 20 in $\mathcal{O}(10)$ nights.}
    \label{fig:highspec}
\end{figure}

The clear detection of Ba-enhanced MS+WD systems despite the relatively low SNR of our spectra underscores the strength of the signal and motivates deeper follow-up observations. Higher-SNR ($\gtrsim 20$) spectra, even at lower resolving power, would enable precise abundance determinations for individual systems, and hence constrain the efficiency of mass transfer as a function of orbital parameters, WD mass and metallicity.

We propose a survey of $\sim 1000$ MS+WD systems from the catalog of \citet{Shahaf_2024_TriageGaia} using the HighSpec spectrograph \citep{SoferRimalt_2024_HighSpecHighResolutiona} on the upcoming Multi-Aperture Spectroscopic Telescope (MAST; Ben-Ami et al., in prep) array at Neot Smadar, Israel, with a particular focus on systems exhibiting sub-solar metallicities and moderate eccentricities. HighSpec employs a custom narrow bandpass ion-etched grating at $R \sim 20{,}000$ and achieves an end-to-end efficiency of 55\%, making it well suited for observing prominent spectral lines even in relatively faint targets. Our primary objective will be to measure the strong Ba\,II absorption line at 4555\,\AA. The surrounding $\sim 120$\,\AA\ band also includes numerous relatively strong Fe lines, enabling precise determination of [Fe/H]. Figure~\ref{fig:highspec} shows the limiting magnitude achievable at various SNRs as a function of exposure time, overlaid on the JKC $B$-band cumulative magnitude distribution of MS+WD systems with sub-solar metallicities (as per \citealt{Andrae_2023_RobustDatadriven}) accessible from Israel. With up to 20 MAST telescopes operating simultaneously with HighSpec, we can observe $\sim 1000$ of the brightest targets (including roughly half with $e > 0.2$) at SNR > 20 in $\mathcal{O}(10)$ nights.

\begin{acknowledgments}
We thank Ana Escorza and Amaya Sinha for useful discussions regarding the abundance analysis.

SS, JMH and HWR acknowledge support from the European Research Council for the ERC Advanced Grant [101054731]. SS was also supported by a Benoziyo prize postdoctoral fellowship.
SBA acknowledges support from the Israel Ministry of Science grant IMOS 715060. SBA's research is supported by the Peter and Patricia Gruber Award; the Andr\'e Deloro Institute for Advanced Research in Space and Optics; the Willner Family Leadership Institute for the Weizmann Institute of Science; and the Israel Science Foundation grant ISF 714022-02. SBA is the incumbent of the Aryeh and Ido Dissentshik Career Development Chair.
ST acknowledges support from the Netherlands Research Council NWO (grant VIDI 203.061).

This work has made use of data from the European Space Agency (ESA) mission Gaia (\url{https://www.cosmos.esa.int/gaia}), processed by the Gaia Data Processing and Analysis Consortium (DPAC; \url{https://www.cosmos.esa.int/web/gaia/dpac/consortium}). Funding for the DPAC has been provided by national institutions, in particular the institutions participating in the Gaia Multilateral Agreement.

Based on observations collected at the European Southern Observatory under ESO programme 114.27T5.001.
\end{acknowledgments}

\facilities{Gaia, Max Planck:2.2m(FEROS)}

\software{\noun{astroplan} \citep{Morris_2018_AstroplanOpen}, \noun{pysme} \citep{Wehrhahn_2023_PySMESpectroscopy}, \noun{pyastronomy} \citep{Czesla_2025_SczeslaPyAstronomy}, \noun{astropy} \citep{AstropyCollaboration_2013_AstropyCommunity,AstropyCollaboration_2018_AstropyProject,AstropyCollaboration_2022_AstropyProject}, \noun{numpy} \citep{Harris_2020_ArrayProgramming}, \noun{scipy} \citep{Virtanen_2020_SciPy10}, \noun{matplotlib} \citep{Hunter_2007_Matplotlib2D}, \noun{brokenaxes} \citep{Dichter_2025_BendichterBrokenaxes}}

\appendix
\restartappendixnumbering

\section{Determination of RV, stellar parameters and line broadening velocities} \label{app:rv_feh_vbroad}

\begin{figure*}
\centering
\subfloat[]{
    \includegraphics[width=0.7\textwidth]{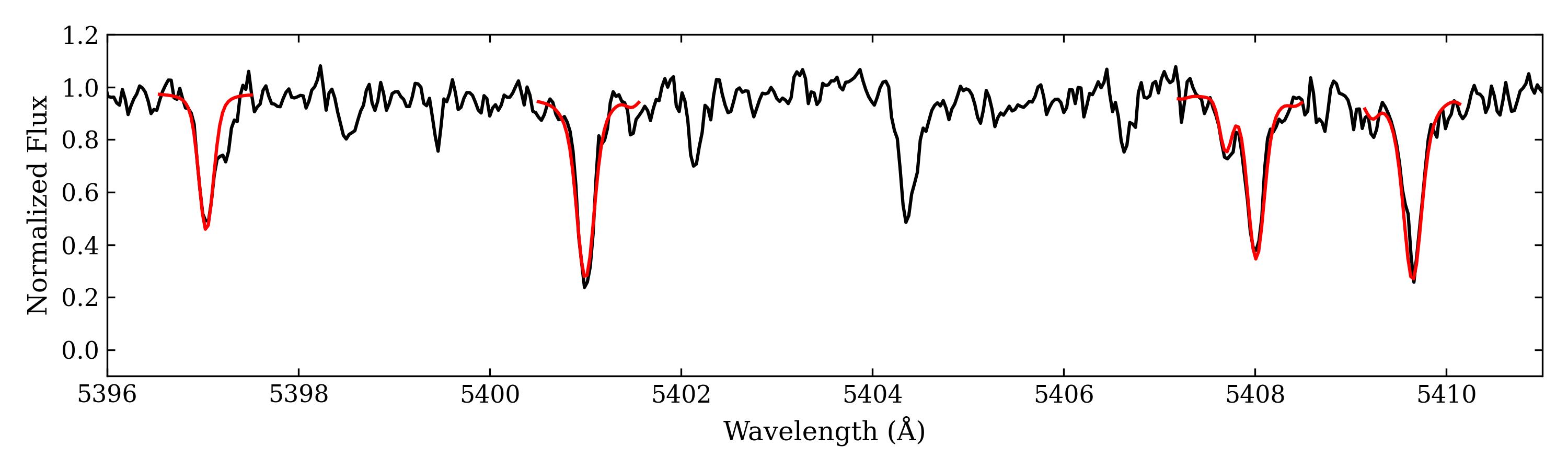}
    \label{fig: eg Fe fit}
}\\
\subfloat[]{
    \includegraphics[width=0.7\textwidth]{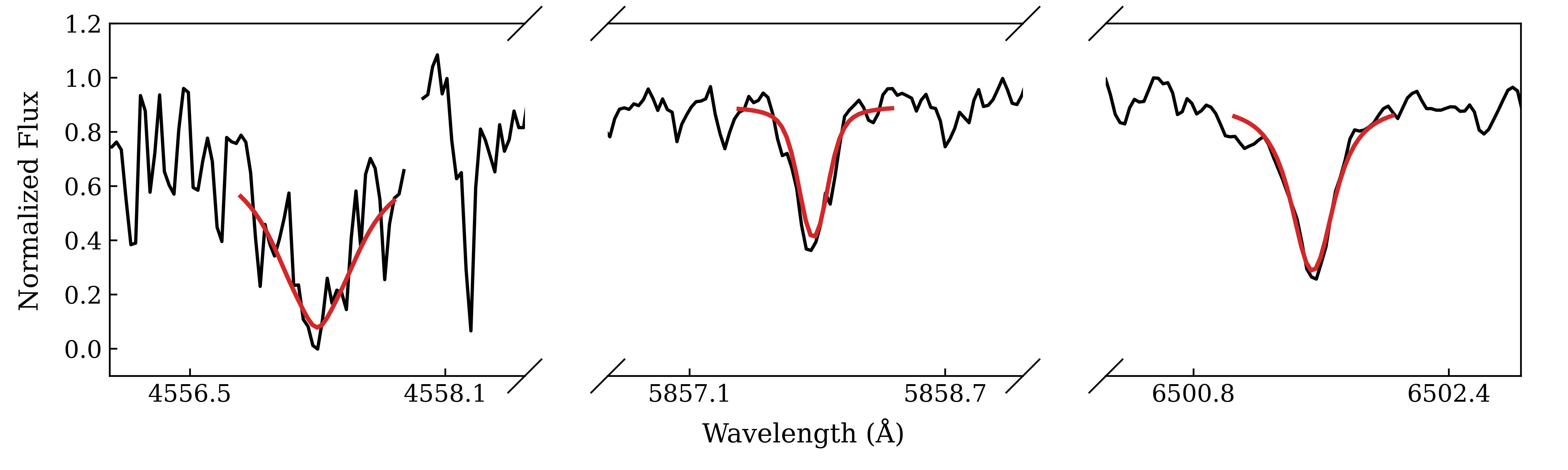}
    \label{fig: eg Ba fit}
}\\
\subfloat[]{
    \includegraphics[width=0.49\textwidth]{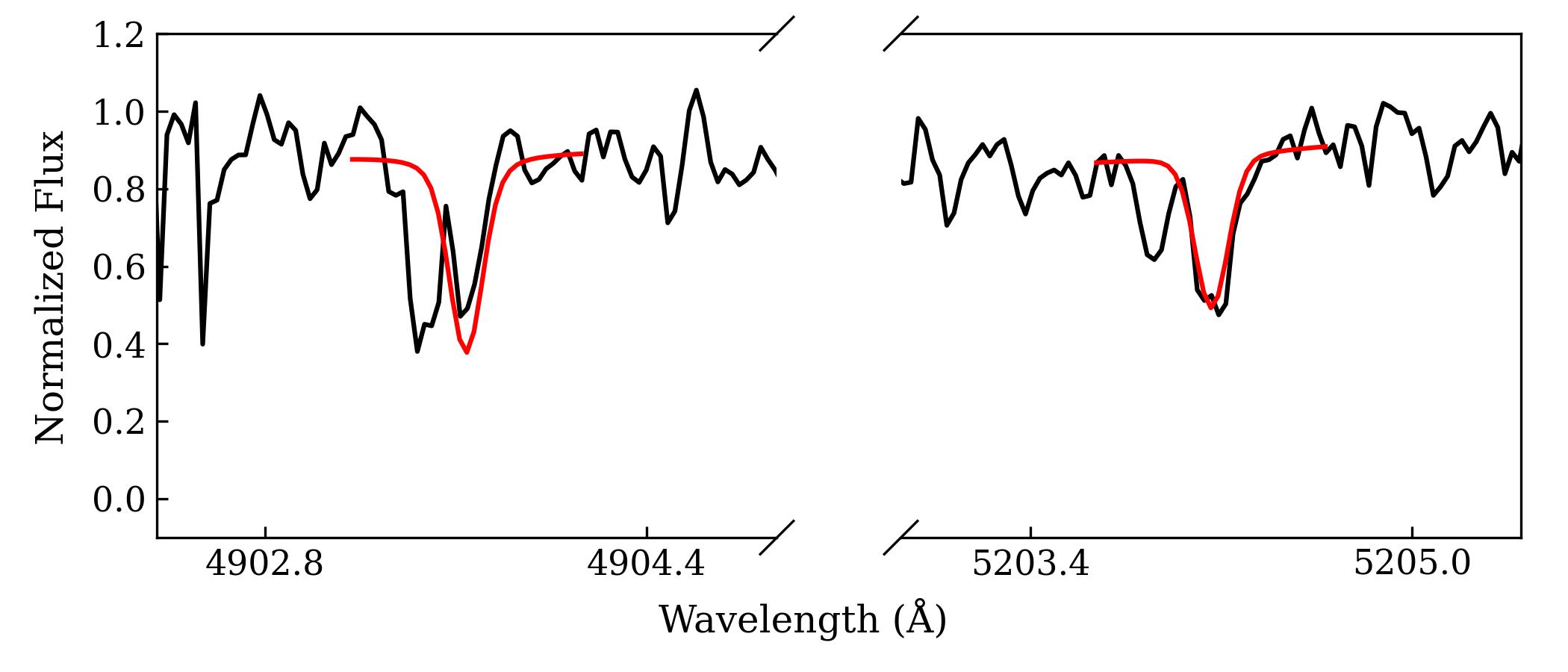}
    \label{fig: eg Y fit}
}\\
\caption{Plots showing \noun{pysme} fits (red) of (a) selected Fe, (b) Ba and (c) Y lines sampled on the wavelength grid of the observed spectrum (black) of Gaia~DR3~3208829887146528384. We obtain fit values of [Fe/H] $= -0.77 \pm 0.06$, [Ba/Fe] $= 1.40 \pm 0.07$ and [Y/Fe] $= 1.51 \pm 0.22$.}
\label{fig:sample fit idx7}
\end{figure*}

The determination of effective temperature ($T_\text{eff}$), Fe abundance ([Fe/H]; here used synonymously with metallicity), microturbulence velocity ($v_\text{mic}$) and rotational broadening velocity ($v_\text{broad}$) was carried out using a set of Fe lines custom-selected for each star.
As the selected Fe lines are minimally pressure-broadened and thus largely insensitive to $\log g$, surface gravity ($\log g$) was fixed to values obtained from Gaia XP–based catalogs \citep{Andrae_2023_RobustDatadriven,Zhang_2023_Parameters220}, which also provided seed values for $T_\text{eff}$ and [Fe/H].
As the low SNR of our spectra makes simultaneous fitting of projected rotational velocity ($v \sin i$) and macroturbulence velocity ($v_\text{mac}$) challenging, we followed \citet{Buder_2021_GALAHSurvey} and fit for $v_\text{broad}$ (defined as $v \sin i$ with $v_\text{mac}$ set to 0) instead. 

For each selected Fe line, we extracted 1\,\AA-wide spectral segments centered on the line, merging overlapping segments. Lines were chosen based on the following criteria: (1) the SNR in the line region exceeded 10, except for three stars where the threshold was lowered to 5 due to fewer than five lines meeting the SNR $>10$ condition; (2) the Fe line depth would exceed 0.4 at [Fe/H] = 0 for the star's $T_\text{eff}$ and $\log g$; (3) there were no lines of other species within $\pm 0.5$\,\AA\ exceeding 20\% of the Fe line's depth; and (4) the mean value of the continuum trend (see next) across the $3\,$\AA\ segment was greater than 0.9, with a maximum variation of less than 0.02. The fourth condition excluded regions where the blaze function caused the pipeline's continuum normalization quality, as well as the local SNR, to drop considerably. To reduce computational expenses, when the number of such segments exceeded 30 (for 5 spectra), we selected the 30 segments with the deepest lines. 

To correct for the local continuum trend, we fit a smoothing spline to continuum points in 60\,\AA\ windows centered on each segment of interest using the \noun{scipy} function \code{scipy.interpolate.make\_smoothing\_spline}. For this procedure, synthetic spectra were created with stellar parameters as described above and Doppler shifted to stellar rest frames using a nominal radial velocity (RV) determined using the H$\alpha$ line. Continuum points were denoted as those where the model flux exceeded 0.98, which ensured a sufficient number of points for robust spline fits with negligible bias from absorption features.

The relative uncertainty in the spectrum in each segment was estimated from the standard deviation of the continuum-identified points in a 20\,\AA\ window centered on the segment. To isolate the noise estimate from the continuum trend, the normalized flux in the windows was first divided by the continuum spline. 

To determine the local RV value around each line segment, we used a 20\,\AA-wide region centered on the corresponding 1\,\AA\ segment. Regions (and corresponding 1\,\AA\ segments) with more than 25\% of the flux points masked were discarded to ensure reliability both in RV estimation and continuum de-trending. Synthetic spectra were generated using the given stellar parameters, including only lines with absorption depths greater than 0.1. The RV was determined using cross-correlation with a built-in method from \noun{pymse}. A $2\sigma$ clipping procedure was then used to identify and discard any windows with unreliable RVs, typically caused by an insufficient number of deep lines in the region. 
Across all spectra, the mean (max) range in the measured RVs across all usable windows is $0.93\, (1.7)$\,km\,s$^{-1}$, and the mean (max) standard deviation is $0.26\, (0.54)$\,km\,s$^{-1}$. 

Each 1\,\AA\ segment was then Doppler-shifted to Earth's rest frame using its locally determined RV value, rather than a global mean, to avoid potential inter- and intra-order wavelength calibration drifts. Subsequently, segments with any masked data points within the core line region --- defined as a $\pm 0.2$\,\AA\ interval around the line center --- were removed from further analysis. After this final filtering step, between 6 and 30 segments per spectrum remained, with a mean of 21 segments per spectrum.

Finally, we conducted a Markov Chain Monte Carlo (MCMC) analysis to fit for $T_\text{eff}$, [Fe/H], $v_\text{mic}$ and $v_\text{broad}$, treating all segments simultaneously. For each star (and its corresponding fixed $\log g$ value), we first created a bank of synthetic spectra using \noun{pysme} on a grid of $T_\text{eff}$, [Fe/H], and $v_\text{mic}$ values at steps of 100,K, 0.05,dex, and 0.1,km,s$^{-1}$, respectively, and with a wavelength resolution of 0.003,\AA. The line list passed to \noun{pysme} during this synthesis was restricted to Fe lines in the selected segments to avoid prohibitively long computation times, as runtime scaled linearly with the number of lines; this restriction does not affect our results, since the Fe lines were previously chosen to avoid significant blending with lines from other species. During each MCMC iteration, unbroadened synthetic spectra were interpolated from the precomputed grid and rotationally and instrumentally broadened using \code{fastRotBroad} (with a limb-darkening parameter of 0.81) and \code{instrBroadGaussFast} from \noun{pyastronomy} \citep[version 0.22.0; ][]{Czesla_2025_SczeslaPyAstronomy}. The instrumental resolution at the center of each segment was determined from the wavelength solution, assuming FEROS’s sampling of 2.2 pixels per resolution element \citep{Kaufer_1998_FEROSNew}. The resulting broadened spectra were then interpolated onto the FEROS wavelength grid, multiplied by the continuum trend, and compared to the observed spectra using the standard log-likelihood function ($-0.5\chi^2$) computed from all data points across the selected Fe segments. Comparison of spectra broadened with \noun{pysme} and \noun{pyastronomy} confirmed that their maximum deviation was always below 1\%. The MCMC posteriors were unimodal but occasionally asymmetric, especially for $v_\text{mic}$, and therefore we report the maximum-likelihood estimate (MLE) for each parameter, with the 68th percentile interval around the MLE serving as the uncertainty estimate. 
Lastly, we tested the insensitivity of the fit to $\log g$ by refitting [Fe/H] as a function of $\log g$ with the other parameters held fixed to their best-fit values, confirming that [Fe/H] remains stable within the estimated uncertainties.

Ba and Y abundances were measured using an analogous MCMC procedure, fixing the stellar parameters to the previously determined values. An example of the spectral fit to Fe, Ba and Y lines is shown in Figure \ref{fig:sample fit idx7}.

\onecolumngrid
\section{Additional Plots} \label{app:addn plots}

\begin{figure*}[!htbp]
\centering
\begin{minipage}[t]{0.48\textwidth}
    \centering
    \includegraphics[width=\linewidth]{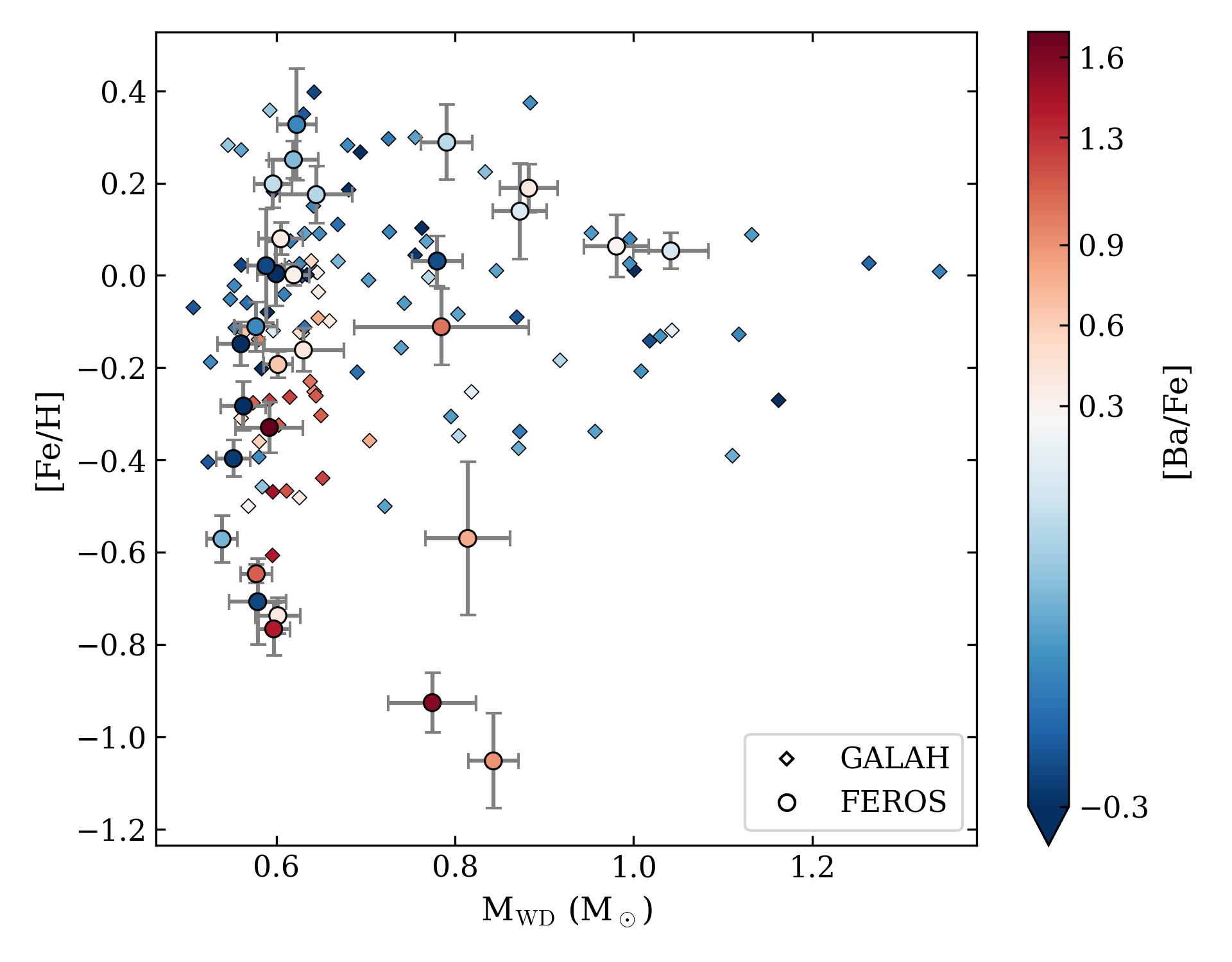}
    \caption{[Ba/Fe] as a function of [Fe/H] and WD mass.}
    \label{fig:ba feh mwd}
\end{minipage}%
\hspace{0.02\textwidth} 
\begin{minipage}[t]{0.48\textwidth}
    \centering
    \includegraphics[width=\linewidth]{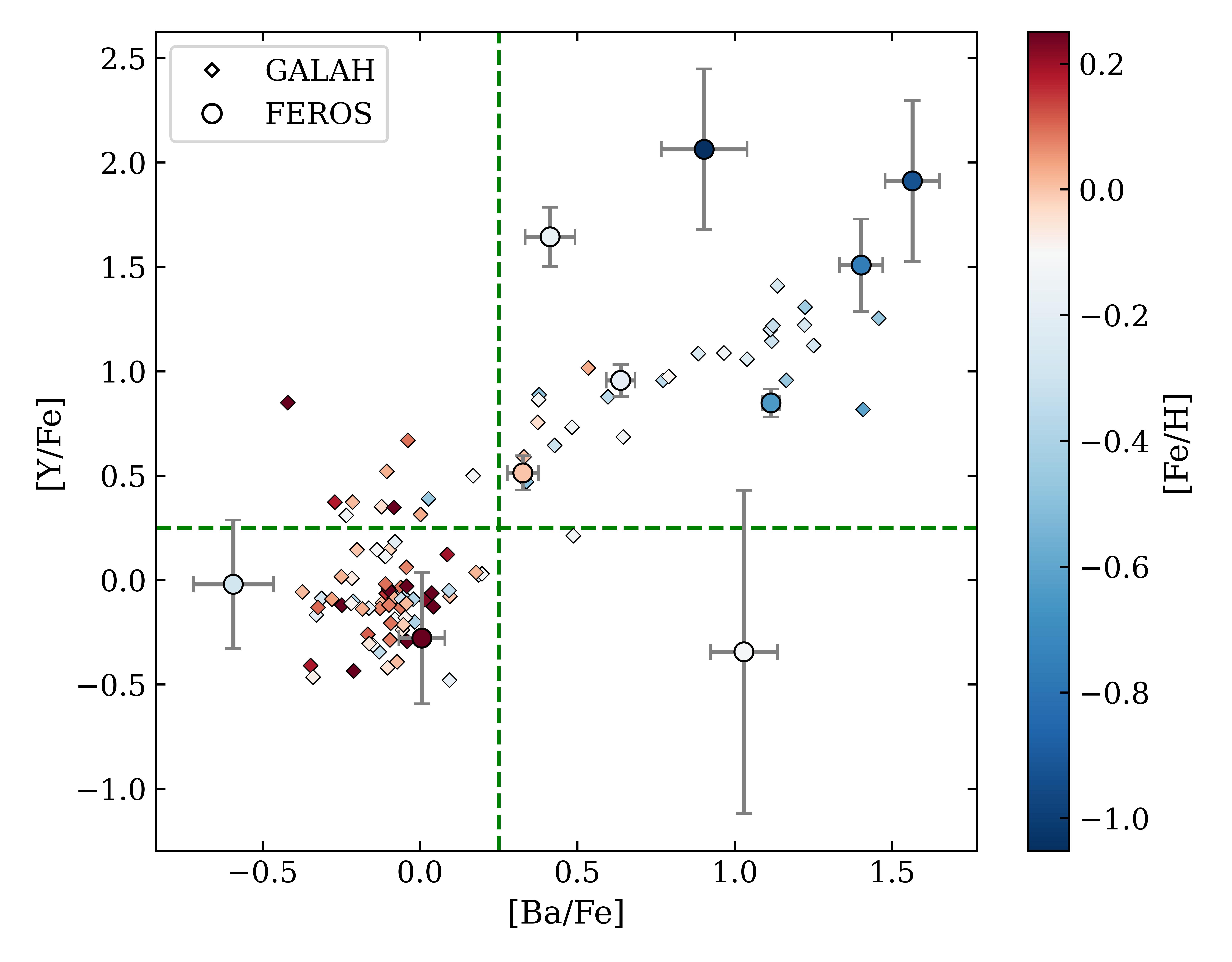}
    \caption{[Ba/Fe] vs [Y/Fe] coloured by [Fe/H], for the 10 targets having Y measurements. The green dashed line denotes the enrichment boundary. Ba-enhanced stars generally exhibit Y enhancement, reducing the likelihood of spurious Ba/Y enhancement.}
    \label{fig:ba y comp}
\end{minipage}
\end{figure*}

\begin{figure}[!htbp]
\centering
\subfloat{
    \includegraphics[width=0.5\columnwidth]{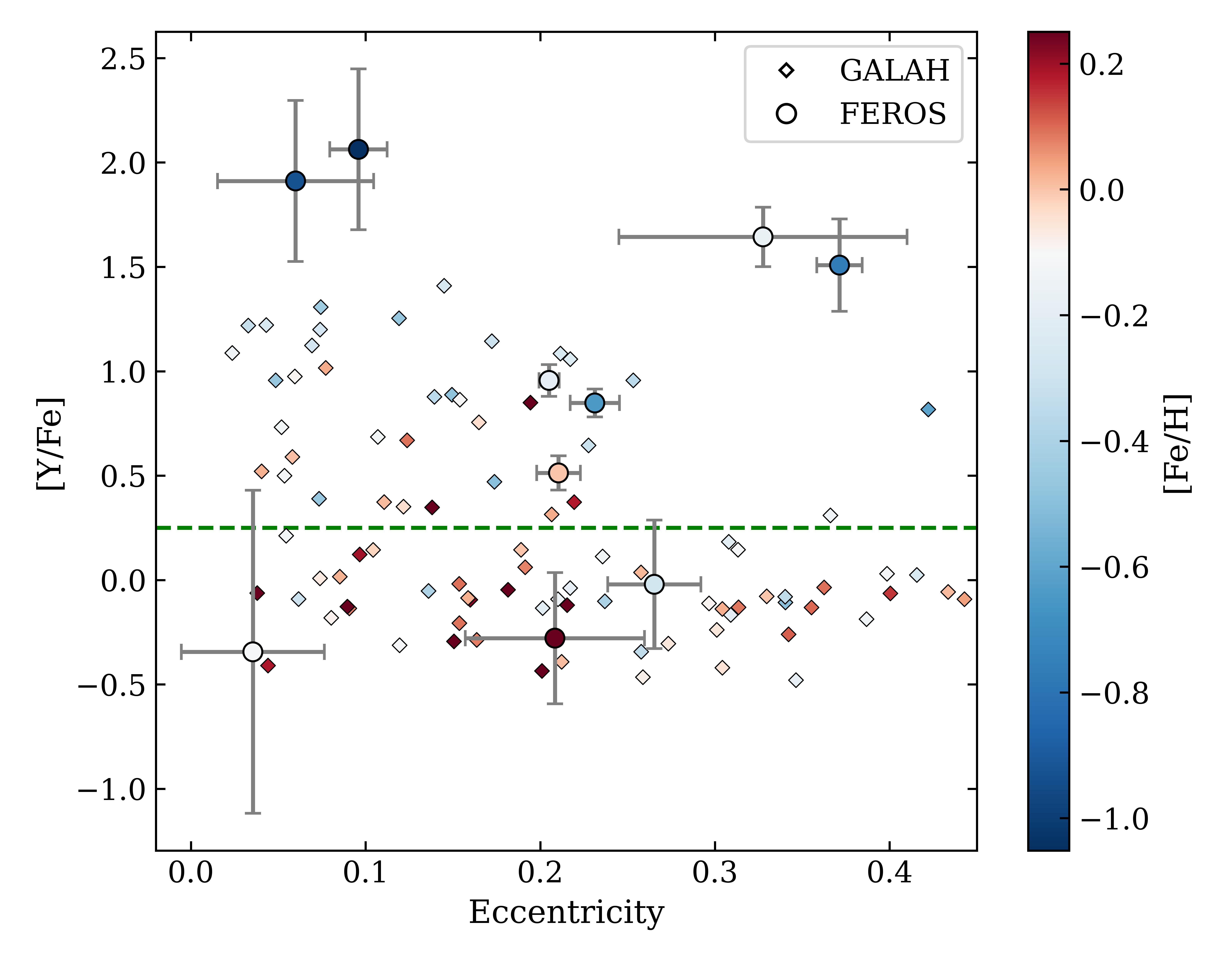}
    \label{fig: Y ecc}
}
\subfloat{
    \includegraphics[width=0.5\columnwidth]{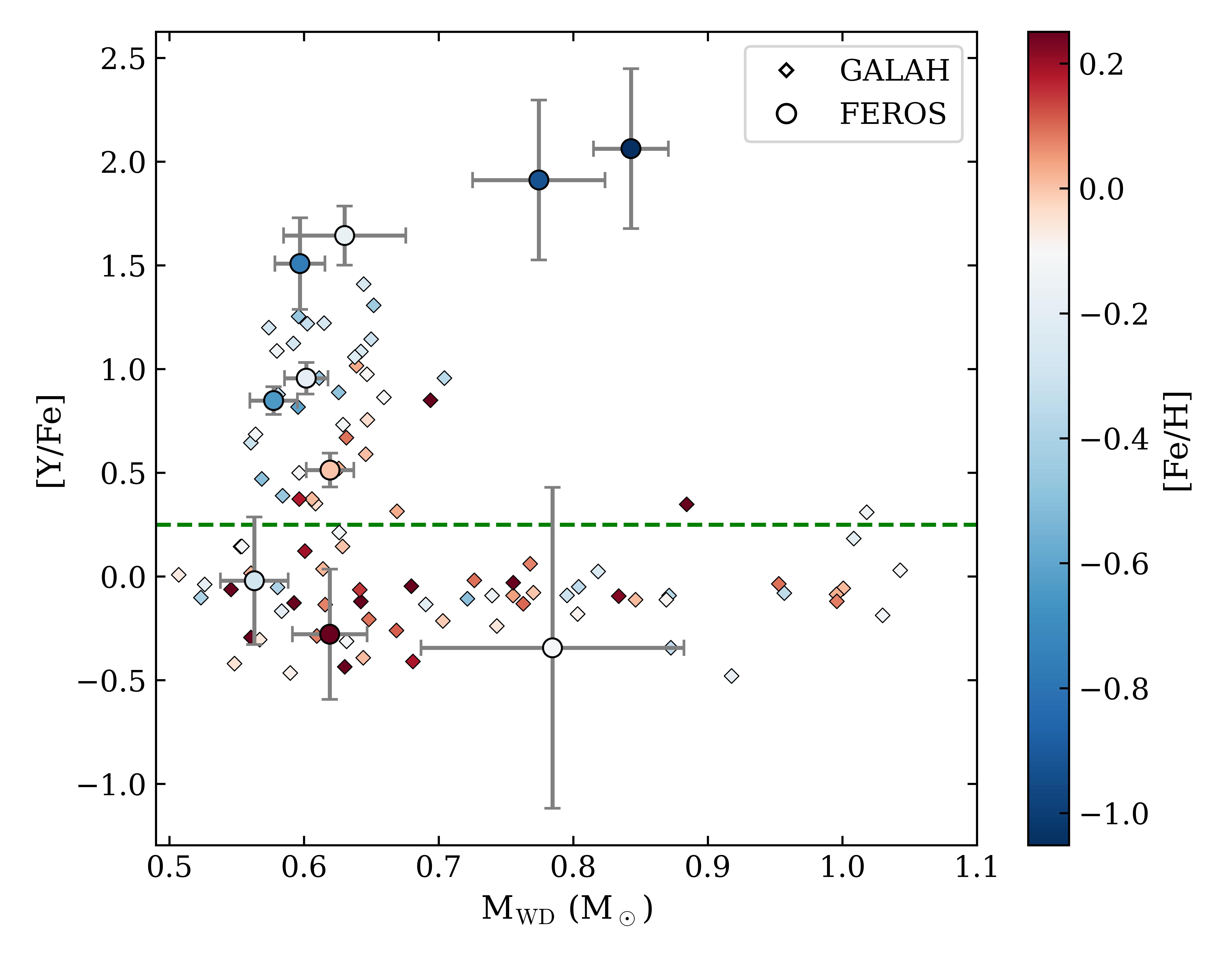}
    \label{fig: Y mwd}
}
\caption{[Y/Fe] as function of (a) eccentricity and (b) \mwd.}
\label{fig: Y color plots}
\end{figure}

Figure \ref{fig:ba feh mwd} shows Ba abundance as a function of [Fe/H] and \mwd\ as a complementary plot to figure 3 of \citetalias{Rekhi_2024_BaEnrichment}. Y abundance plots that complement Figure \ref{fig: Ba color plots} are presented in Figures \ref{fig:ba y comp} and \ref{fig: Y color plots}.

\section{Data Tables}

\begin{deluxetable}{cccccccccccc}
\tabletypesize{\footnotesize}
\tablecaption{List of systems analyzed in this work, showing measured abundances along with orbital and stellar parameters and component masses. Periods and eccentricities originate from \citet{GaiaCollaboration_2023_NSS}, $\log(g)$ values are taken from \citet{Andrae_2023_RobustDatadriven}, or if absent there, from \citet{Zhang_2023_Parameters220}. \m1 and \mwd\ are obtained from \citet{GaiaCollaboration_2023_NSS} and \citet{Shahaf_2023_TriageGaia} respectively.\label{tab:data_table}}
\tablehead{
\colhead{Gaia DR3 ID} &
\colhead{Period (d)} &
\colhead{Eccentricity} &
\colhead{\mwd\ (\msun)} &
\colhead{\m1\ (\msun)} &
\colhead{[Fe/H]} &
\colhead{[Ba/Fe]} &
\colhead{[Y/Fe]} &
\colhead{T$_\text{eff}$} &
\colhead{$\log g$} &
\colhead{$v_\text{mic}$} &
\colhead{$v_\text{broad}$}
}
\startdata
3036788173176962048 & $195.6 \pm 0.6$ & $0.27 \pm 0.05$ & $0.59 \pm 0.04$ & $0.97 \pm 0.07$ & $-0.33 \pm 0.06$ & $1.69 \pm 0.07$ & -- & $6041 \pm 56$ & $4.47$ & $1.12 \pm 0.11$ & $5.52 \pm 0.22$  \\
4790658375100436736 & $939.2 \pm 35.8$ & $0.06 \pm 0.04$ & $0.77 \pm 0.05$ & $1.15 \pm 0.06$ & $-0.93 \pm 0.06$ & $1.56 \pm 0.09$ & $1.91 \pm 0.39$ & $5793 \pm 61$ & $4.32$ & $1.55 \pm 0.12$ & $4.79 \pm 0.28$  \\
3208829887146528384 & $684.5 \pm 2.5$ & $0.37 \pm 0.01$ & $0.60 \pm 0.02$ & $0.88 \pm 0.05$ & $-0.77 \pm 0.06$ & $1.40 \pm 0.07$ & $1.51 \pm 0.22$ & $5439 \pm 39$ & $4.48$ & $0.55 \pm 0.17$ & $3.93 \pm 0.26$  \\
5620705644518714752 & $239.2 \pm 0.2$ & $0.23 \pm 0.01$ & $0.58 \pm 0.02$ & $0.86 \pm 0.05$ & $-0.65 \pm 0.02$ & $1.12 \pm 0.03$ & $0.85 \pm 0.07$ & $5405 \pm 13$ & $4.48$ & $0.54 \pm 0.08$ & $3.72 \pm 0.16$  \\
6091100873778091392 & $987.3 \pm 65.1$ & $0.04 \pm 0.04$ & $0.78 \pm 0.10$ & $1.04 \pm 0.09$ & $-0.11 \pm 0.08$ & $1.03 \pm 0.11$ & $-0.34 \pm 0.77$ & $5824 \pm 67$ & $4.33$ & $1.12 \pm 0.14$ & $3.91 \pm 0.38$  \\
5406806102703951360 & $832.4 \pm 7.7$ & $0.10 \pm 0.02$ & $0.84 \pm 0.03$ & $1.02 \pm 0.06$ & $-1.05 \pm 0.10$ & $0.90 \pm 0.14$ & $2.06 \pm 0.39$ & $5320 \pm 102$ & $4.48$ & $0.82 \pm 0.35$ & $4.50 \pm 0.57$  \\
3772123820203876736 & $529.2 \pm 6.4$ & $0.07 \pm 0.08$ & $0.81 \pm 0.05$ & $1.04 \pm 0.06$ & $-0.57 \pm 0.17$ & $0.79 \pm 0.24$ & -- & $6007 \pm 157$ & $4.41$ & $1.45 \pm 0.35$ & $4.34 \pm 0.92$  \\
5397347794427206656 & $720.7 \pm 1.0$ & $0.21 \pm 0.01$ & $0.60 \pm 0.02$ & $0.95 \pm 0.05$ & $-0.19 \pm 0.03$ & $0.64 \pm 0.05$ & $0.96 \pm 0.08$ & $5742 \pm 22$ & $4.48$ & $1.08 \pm 0.05$ & $4.75 \pm 0.15$  \\
6181668466886465280 & $213.6 \pm 0.8$ & $0.33 \pm 0.08$ & $0.63 \pm 0.05$ & $1.03 \pm 0.06$ & $-0.16 \pm 0.05$ & $0.41 \pm 0.08$ & $1.64 \pm 0.14$ & $5854 \pm 35$ & $4.39$ & $1.27 \pm 0.10$ & $11.23 \pm 0.20$  \\
5722712938857848192 & $614.8 \pm 3.5$ & $0.05 \pm 0.04$ & $0.88 \pm 0.03$ & $1.19 \pm 0.06$ & $0.19 \pm 0.05$ & $0.40 \pm 0.10$ & -- & $6347 \pm 52$ & $4.30$ & $0.95 \pm 0.14$ & $9.10 \pm 0.21$  \\
2914005812780342272 & $399.2 \pm 1.5$ & $0.32 \pm 0.02$ & $0.60 \pm 0.03$ & $0.93 \pm 0.06$ & $-0.74 \pm 0.04$ & $0.39 \pm 0.07$ & -- & $5479 \pm 27$ & $4.36$ & $0.82 \pm 0.09$ & $4.18 \pm 0.19$  \\
5164360455867611264 & $922.0 \pm 29.9$ & $0.26 \pm 0.03$ & $0.60 \pm 0.02$ & $0.84 \pm 0.06$ & $0.08 \pm 0.03$ & $0.37 \pm 0.08$ & -- & $5019 \pm 17$ & $4.51$ & $0.00 \pm 0.16$ & $3.81 \pm 0.32$  \\
4795585179328518016 & $414.7 \pm 0.6$ & $0.21 \pm 0.01$ & $0.62 \pm 0.02$ & $1.00 \pm 0.05$ & $0.00 \pm 0.02$ & $0.33 \pm 0.05$ & $0.51 \pm 0.08$ & $5860 \pm 16$ & $4.49$ & $0.94 \pm 0.06$ & $5.27 \pm 0.14$  \\
5550946678313327744 & $800.1 \pm 22.2$ & $0.05 \pm 0.02$ & $0.98 \pm 0.04$ & $1.03 \pm 0.06$ & $0.06 \pm 0.07$ & $0.29 \pm 0.13$ & -- & $5851 \pm 51$ & $4.33$ & $0.31 \pm 0.25$ & $7.99 \pm 0.30$  \\
2923820637686349568 & $898.7 \pm 24.2$ & $0.07 \pm 0.04$ & $0.87 \pm 0.03$ & $1.18 \pm 0.06$ & $0.14 \pm 0.10$ & $0.16 \pm 0.17$ & -- & $6061 \pm 87$ & $4.25$ & $0.84 \pm 0.27$ & $8.44 \pm 0.39$  \\
5327240562221030528 & $162.4 \pm 0.2$ & $0.09 \pm 0.02$ & $1.04 \pm 0.04$ & $0.87 \pm 0.06$ & $0.05 \pm 0.04$ & $0.15 \pm 0.09$ & -- & $5359 \pm 23$ & $4.62$ & $0.71 \pm 0.10$ & $4.46 \pm 0.25$  \\
5329433194583181952 & $546.2 \pm 2.2$ & $0.29 \pm 0.02$ & $0.60 \pm 0.02$ & $0.91 \pm 0.06$ & $0.20 \pm 0.05$ & $0.11 \pm 0.12$ & -- & $5561 \pm 32$ & $4.51$ & $0.15 \pm 0.21$ & $4.49 \pm 0.32$  \\
2964151136511557120 & $800.4 \pm 14.6$ & $0.04 \pm 0.02$ & $0.79 \pm 0.03$ & $1.23 \pm 0.06$ & $0.29 \pm 0.08$ & $0.10 \pm 0.22$ & -- & $6689 \pm 111$ & $4.30$ & $1.22 \pm 0.27$ & $11.94 \pm 0.39$  \\
6169594184243050752 & $821.9 \pm 37.8$ & $0.25 \pm 0.06$ & $0.64 \pm 0.04$ & $0.95 \pm 0.06$ & $0.18 \pm 0.06$ & $0.09 \pm 0.13$ & -- & $5659 \pm 41$ & $4.46$ & $0.98 \pm 0.12$ & $5.47 \pm 0.28$  \\
6300835156308377344 & $208.8 \pm 0.7$ & $0.21 \pm 0.05$ & $0.62 \pm 0.03$ & $0.94 \pm 0.06$ & $0.25 \pm 0.04$ & $0.01 \pm 0.07$ & $-0.28 \pm 0.31$ & $5590 \pm 26$ & $4.36$ & $0.65 \pm 0.11$ & $4.81 \pm 0.21$  \\
5317828124219671168 & $602.4 \pm 1.0$ & $0.24 \pm 0.01$ & $0.54 \pm 0.02$ & $0.81 \pm 0.05$ & $-0.57 \pm 0.05$ & $-0.00 \pm 0.09$ & -- & $4858 \pm 30$ & $4.56$ & $0.90 \pm 0.17$ & $5.05 \pm 0.32$  \\
2919286664051008640 & $155.2 \pm 0.2$ & $0.30 \pm 0.04$ & $0.58 \pm 0.02$ & $0.92 \pm 0.06$ & $-0.11 \pm 0.05$ & $-0.11 \pm 0.12$ & -- & $5357 \pm 46$ & $4.51$ & $0.88 \pm 0.14$ & $3.97 \pm 0.36$  \\
5744310130364616576 & $906.2 \pm 18.1$ & $0.25 \pm 0.03$ & $0.62 \pm 0.02$ & $0.88 \pm 0.06$ & $0.33 \pm 0.12$ & $-0.11 \pm 0.22$ & -- & $5251 \pm 85$ & $4.45$ & $0.00 \pm 0.33$ & $3.86 \pm 0.64$  \\
5890631694717361408 & $934.4 \pm 30.7$ & $0.05 \pm 0.03$ & $0.78 \pm 0.03$ & $1.07 \pm 0.06$ & $0.03 \pm 0.05$ & $-0.24 \pm 0.13$ & -- & $6001 \pm 44$ & $4.38$ & $0.97 \pm 0.11$ & $6.71 \pm 0.21$  \\
3697174751104247808 & $914.5 \pm 5.5$ & $0.38 \pm 0.04$ & $0.58 \pm 0.03$ & $0.88 \pm 0.07$ & $-0.71 \pm 0.09$ & $-0.25 \pm 0.25$ & -- & $5168 \pm 74$ & $4.54$ & $0.72 \pm 0.41$ & $4.54 \pm 0.61$  \\
6192821814055372800 & $750.6 \pm 6.4$ & $0.32 \pm 0.03$ & $0.59 \pm 0.02$ & $0.88 \pm 0.06$ & $0.02 \pm 0.12$ & $-0.25 \pm 0.23$ & -- & $4959 \pm 123$ & $3.88$ & $0.92 \pm 0.25$ & $4.61 \pm 0.46$  \\
3147042801161881472 & $509.8 \pm 1.7$ & $0.28 \pm 0.02$ & $0.55 \pm 0.02$ & $0.76 \pm 0.05$ & $-0.40 \pm 0.04$ & $-0.27 \pm 0.21$ & -- & $4892 \pm 20$ & $4.57$ & $0.02 \pm 0.12$ & $7.93 \pm 0.44$  \\
2923409729576118912 & $894.9 \pm 19.1$ & $0.24 \pm 0.02$ & $0.60 \pm 0.02$ & $0.93 \pm 0.06$ & $0.00 \pm 0.07$ & $-0.37 \pm 0.16$ & -- & $5392 \pm 40$ & $4.45$ & $0.71 \pm 0.22$ & $3.54 \pm 0.51$  \\
5091887624392418816 & $229.7 \pm 0.5$ & $0.20 \pm 0.04$ & $0.56 \pm 0.03$ & $0.92 \pm 0.07$ & $-0.15 \pm 0.05$ & $-0.52 \pm 0.09$ & -- & $5251 \pm 35$ & $4.51$ & $1.06 \pm 0.11$ & $6.45 \pm 0.23$  \\
3633619813626022784 & $551.2 \pm 3.0$ & $0.27 \pm 0.03$ & $0.56 \pm 0.03$ & $0.92 \pm 0.07$ & $-0.28 \pm 0.05$ & $-0.59 \pm 0.13$ & $-0.02 \pm 0.31$ & $5646 \pm 41$ & $4.25$ & $0.58 \pm 0.20$ & $4.07 \pm 0.30$  \\
\enddata
\tablecomments{This table is also available in machine-readable format in the arXiv ancillary material with uncertainties provided in separate columns. The quoted error for \m1 represents the mean of the upper and lower uncertainties. Quantities above are rounded to 2 decimal digits for readability.}
\end{deluxetable}

\begin{deluxetable}{ccccccccc}
\tablecaption{List of Ba and Y enriched systems from \citetalias{Rekhi_2024_BaEnrichment}, showing GALAH-measured abundances \citep{Buder_2021_GALAHSurvey} along with orbital parameters and component masses. Periods and eccentricities originate from \citet{GaiaCollaboration_2023_NSS}, \m1\ from \citet{Buder_2021_GALAHSurvey} and \mwd\ is derived using Eqn. 2 of \citet{Shahaf_2023_TriageGaia}. \label{tab:r24_data_table}}
\tablehead{
\colhead{Gaia DR3 ID} &
\colhead{GALAH DR3 ID} &
\colhead{Period (d)} &
\colhead{Eccentricity} &
\colhead{\mwd\ (\msun)} &
\colhead{\m1\ (\msun)} &
\colhead{[Fe/H]} &
\colhead{[Ba/Fe]} &
\colhead{[Y/Fe]}
}
\startdata
6472197475176313472 & 170514003301152 & $253.3 \pm 1.0$ & $0.12 \pm 0.07$ & $0.60 \pm 0.03$ & $0.90 \pm 0.04$ & $-0.47 \pm 0.07$ & $1.46 \pm 0.07$ & $1.25 \pm 0.10$  \\
6645994082722928768 & 170712004201185 & $705.2 \pm 5.1$ & $0.42 \pm 0.06$ & $0.60 \pm 0.03$ & $0.89 \pm 0.04$ & $-0.61 \pm 0.08$ & $1.41 \pm 0.07$ & $0.82 \pm 0.10$  \\
4878036323641744384 & 150112002501291 & $846.7 \pm 12.5$ & $0.07 \pm 0.02$ & $0.59 \pm 0.02$ & $0.94 \pm 0.06$ & $-0.27 \pm 0.11$ & $1.25 \pm 0.14$ & $1.12 \pm 0.21$  \\
4646881493008180352 & 140809004901046 & $669.6 \pm 8.7$ & $0.07 \pm 0.07$ & $0.65 \pm 0.03$ & $1.04 \pm 0.05$ & $-0.44 \pm 0.12$ & $1.22 \pm 0.11$ & $1.31 \pm 0.17$  \\
4774266821393569792 & 161218003101346 & $454.2 \pm 1.5$ & $0.04 \pm 0.02$ & $0.61 \pm 0.02$ & $0.98 \pm 0.04$ & $-0.26 \pm 0.06$ & $1.22 \pm 0.08$ & $1.22 \pm 0.11$  \\
6489348928855798272 & 170911004201207 & $436.6 \pm 3.0$ & $0.05 \pm 0.04$ & $0.61 \pm 0.04$ & $0.95 \pm 0.06$ & $-0.47 \pm 0.13$ & $1.16 \pm 0.14$ & $0.96 \pm 0.22$  \\
5495572631335262208 & 171205004601232 & $302.6 \pm 1.7$ & $0.14 \pm 0.06$ & $0.64 \pm 0.03$ & $1.06 \pm 0.06$ & $-0.26 \pm 0.12$ & $1.14 \pm 0.14$ & $1.41 \pm 0.21$  \\
5707071630039205248 & 170417002601038 & $249.2 \pm 0.2$ & $0.03 \pm 0.01$ & $0.60 \pm 0.01$ & $0.93 \pm 0.03$ & $-0.32 \pm 0.04$ & $1.12 \pm 0.04$ & $1.22 \pm 0.05$  \\
4715941990147018752 & 171001002901085 & $206.8 \pm 0.9$ & $0.17 \pm 0.08$ & $0.65 \pm 0.04$ & $0.99 \pm 0.04$ & $-0.30 \pm 0.07$ & $1.12 \pm 0.08$ & $1.14 \pm 0.11$  \\
4881348087024892416 & 171106003601092 & $797.0 \pm 14.9$ & $0.07 \pm 0.03$ & $0.57 \pm 0.01$ & $0.84 \pm 0.03$ & $-0.28 \pm 0.07$ & $1.11 \pm 0.08$ & $1.20 \pm 0.11$  \\
5363208336349003648 & 140311007101125 & $919.5 \pm 10.1$ & $0.22 \pm 0.02$ & $0.64 \pm 0.01$ & $0.97 \pm 0.04$ & $-0.23 \pm 0.06$ & $1.04 \pm 0.06$ & $1.06 \pm 0.09$  \\
2436150255390344320 & 161105003601033 & $855.5 \pm 4.7$ & $0.02 \pm 0.02$ & $0.58 \pm 0.01$ & $0.92 \pm 0.03$ & $-0.14 \pm 0.07$ & $0.97 \pm 0.09$ & $1.09 \pm 0.13$  \\
3623765616756619392 & 160602001601256 & $843.8 \pm 9.8$ & $0.21 \pm 0.03$ & $0.64 \pm 0.03$ & $0.95 \pm 0.05$ & $-0.25 \pm 0.11$ & $0.88 \pm 0.15$ & $1.08 \pm 0.24$  \\
6669412928776138112 & 160521004801173 & $285.8 \pm 0.8$ & $0.06 \pm 0.05$ & $0.65 \pm 0.03$ & $1.06 \pm 0.04$ & $-0.09 \pm 0.06$ & $0.79 \pm 0.05$ & $0.97 \pm 0.07$  \\
4693571910684108928 & 140809004901240 & $519.3 \pm 2.9$ & $0.25 \pm 0.03$ & $0.70 \pm 0.02$ & $0.91 \pm 0.03$ & $-0.36 \pm 0.07$ & $0.77 \pm 0.07$ & $0.96 \pm 0.10$  \\
6081919436288801024 & 170508003301353 & $535.3 \pm 1.2$ & $0.11 \pm 0.02$ & $0.56 \pm 0.02$ & $0.88 \pm 0.04$ & $-0.12 \pm 0.10$ & $0.65 \pm 0.15$ & $0.69 \pm 0.22$  \\
5435770125183875200 & 151225003801120 & $414.8 \pm 2.3$ & $0.14 \pm 0.04$ & $0.58 \pm 0.03$ & $0.86 \pm 0.05$ & $-0.36 \pm 0.10$ & $0.60 \pm 0.13$ & $0.88 \pm 0.20$  \\
2600643791975931264 & 140710006102184 & $840.4 \pm 17.7$ & $0.08 \pm 0.07$ & $0.64 \pm 0.03$ & $1.06 \pm 0.05$ & $0.03 \pm 0.08$ & $0.53 \pm 0.13$ & $1.02 \pm 0.19$  \\
2545472433093014912 & 131217001801154 & $884.5 \pm 5.5$ & $0.05 \pm 0.02$ & $0.63 \pm 0.01$ & $0.93 \pm 0.02$ & $-0.13 \pm 0.05$ & $0.48 \pm 0.04$ & $0.73 \pm 0.06$  \\
4650039354829199616 & 161219003601130 & $603.3 \pm 0.9$ & $0.23 \pm 0.01$ & $0.56 \pm 0.01$ & $0.76 \pm 0.02$ & $-0.31 \pm 0.06$ & $0.43 \pm 0.08$ & $0.64 \pm 0.11$  \\
3285029929897944320 & 141104004301120 & $693.7 \pm 9.4$ & $0.15 \pm 0.06$ & $0.63 \pm 0.03$ & $0.93 \pm 0.05$ & $-0.48 \pm 0.11$ & $0.38 \pm 0.12$ & $0.89 \pm 0.19$  \\
4622751301587564416 & 160919004601297 & $177.1 \pm 0.8$ & $0.15 \pm 0.07$ & $0.66 \pm 0.03$ & $0.97 \pm 0.05$ & $-0.10 \pm 0.10$ & $0.38 \pm 0.14$ & $0.86 \pm 0.21$  \\
5371201888966580992 & 160524002101389 & $337.2 \pm 0.7$ & $0.16 \pm 0.02$ & $0.65 \pm 0.02$ & $1.05 \pm 0.05$ & $-0.04 \pm 0.08$ & $0.37 \pm 0.10$ & $0.76 \pm 0.14$  \\
5386048560107201792 & 150408004101125 & $288.8 \pm 0.5$ & $0.17 \pm 0.02$ & $0.57 \pm 0.01$ & $0.80 \pm 0.02$ & $-0.50 \pm 0.05$ & $0.34 \pm 0.04$ & $0.47 \pm 0.06$  \\
6402712399045698560 & 170614005101079 & $734.6 \pm 7.6$ & $0.06 \pm 0.05$ & $0.65 \pm 0.03$ & $1.05 \pm 0.05$ & $0.01 \pm 0.08$ & $0.33 \pm 0.10$ & $0.59 \pm 0.16$  \\
\enddata
\tablecomments{The full table from \citetalias{Rekhi_2024_BaEnrichment} is available in machine-readable format in the arXiv ancillary material with uncertainties provided in separate columns. The quoted error for \m1 represents the mean of the upper and lower uncertainties. Quantities above are rounded to 2 decimal digits for readability.}
\end{deluxetable}

\clearpage
\bibliographystyle{aasjournalv7}
\bibliography{main}

\end{document}